\documentclass[aps,pre,
twocolumn,superscriptaddress,showpacs,floatfix]{revtex4-1}

\usepackage{graphicx}
\usepackage{amssymb,amsmath,amsbsy,amsfonts,bm}
\usepackage{multirow}
\usepackage{tikz}
\usepackage{hyperref}
\usepackage{gensymb}
\usepackage{threeparttable}
\usepackage{csquotes}
\usepackage{braket}
\usepackage{cleveref}
\usepackage[percent]{overpic}
\usepackage{color}

\definecolor{darklavender}{rgb}{0.45, 0.31, 0.59}
\definecolor{amethyst}{rgb}{0.6, 0.4, 0.8}
\definecolor{paulcolour}{rgb}{0.78, 0.082, 0.52}
\definecolor{paulcolour2}{rgb}{0.1, 0.6, 0.8}

\newlength{\myl}
\newcommand{\dif}{\mathrm{d}}

\newcommand{\SUM}[2]{{\setlength{\myl}{\widthof{$\displaystyle\sum_{#1}^{#2}$}*\real{0.5}-\widthof{$\displaystyle\sum$}*\real{0.5}}\sum_{#1}^{#2}\;\hspace{-\the\myl}}}
\newcommand{\INT}[3]{\settowidth{\myl}{$\displaystyle\int_{#1}^{#2}$}{\int_{#1}^{#2}\;\;\;\hspace{-\the\myl}\dif #3}\,}

\usepackage[normalem]{ulem}

\usepackage{pifont}
\newcommand{\circledbullet}{\mathbin{\ooalign{$\circledcirc$\cr\hidewidth$\bullet$\hidewidth}}}

\newcommand{\newcircledcirc}{\mathbin{\ooalign{$\circledcirc$\cr\hidewidth${\color{white}\bullet}$\hidewidth}}}

\newcommand{\myCCC}{{\protect\circledbullet}}
\newcommand{\myCARRIER}{{\protect\newcircledcirc}}
\newcommand{\myCARGO}{\bullet}

\newcommand{\rhomyCARGO}{\rho_\myCARGO^\text{t}}
\newcommand{\rhomyCARRIER}{\rho_\myCARRIER^\text{t}}

\newcommand{\rhoA}{\rho_{\myCARRIER}}
\newcommand{\rhoB}{\rho_{\myCARGO}}

\newcommand{\fu}[2]{\ensuremath{\left\{\begin{array}{c}#1\\#2\end{array}\right.}}
\newcommand{\fuC}[2]{\ensuremath{\left.\begin{array}{l}\mbox{#1}\\\mbox{#2}\end{array}\right.}}

\begin{document}
\title{Statistics of carrier-cargo complexes}

\author{Ren\'e Wittmann}
\email{rene.wittmann@hhu.de}
\affiliation{\mbox{Institut f\"ur Theoretische Physik II: Weiche Materie, Heinrich-Heine-Universit\"at D\"usseldorf, 40225 D\"usseldorf, Germany}}

\author{Paul A. Monderkamp}
\affiliation{\mbox{Institut f\"ur Theoretische Physik II: Weiche Materie, Heinrich-Heine-Universit\"at D\"usseldorf, 40225 D\"usseldorf, Germany}}

\author{Hartmut L\"owen}
\affiliation{\mbox{Institut f\"ur Theoretische Physik II: Weiche Materie, Heinrich-Heine-Universit\"at D\"usseldorf, 40225 D\"usseldorf, Germany}}

\date{\today}

\begin{abstract}
We explore the statistics of assembling soft-matter building blocks to investigate the
uptake and encapsulation of cargo particles by carriers engulfing their load.
While the such carrier-cargo complexes are important for many applications out of equilibrium,
such as drug delivery and synthetic cell encapsulation, we uncover here the basic statistical physics in minimal hard-core-like models for particle uptake.
Introducing an exactly solvable equilibrium model in one dimension, we demonstrate that the formation of carrier-cargo complexes can be largely tuned by both the cargo concentration and the carriers' interior size.
 These findings are intuitively explained by interpreting the internal free space (partition function) of the cargo inside a carrier as its engulfment strength,
which can be mapped to an external control parameter (chemical potential) of an additional effective particle species.
Assuming a hard carrier membrane, such a mapping can be exactly applied to account for multiple cargo uptake involving various carrier or cargo species and even attractive uptake mechanisms,
while soft interactions require certain approximations.
  We further argue that the Boltzmann occupation law identified within our approach is broken when particle uptake is governed by non-equilibrium forces.
  Speculating on alternative occupation laws using effective parameters, we put forward a Bose-Einstein-like phase transition associated with polydisperse carrier properties.
\end{abstract}

\maketitle

\section{Introduction}

When two or more soft-matter
building blocks are combined, self-organization can
lead to novel hierarchical structures with unusual
material properties \cite{thunemann2002polyelectrolyte,hamley2007biological,vogel2015advances,yan2016reconfiguring,harder2018hierarchical,massana2021arrested,wan2021polymerizationcolloidalmolecules,aubret2021metamachines,vankesteren2022printing,hokmabad2022spontaneously}.
One archetypal problem is a mesoscopic carrier particle that
swallows or uptakes smaller cargo particles to build a carrier-cargo complex.
These superstructures can occur in quite diverse situations ranging from greenhouse gases stored in porous liquids \cite{giri2015liquids,avila2021high} over adsorbed or encapsulated drugs which need to be delivered to a target \cite{karimi2016smart,bo2020multifunctional,ma2017mesoporous,alapan2018soft,sun2019coated,perry2020_encapsulation}
to molecules or nanoparticles that penetrate through the membrane of synthetic and living cells  \cite{madani2011mechanisms,ma2013influence,tan2015cell,abels2016introduction,mulla2018shaping,steinkuhler2020controlled,ashraf2020quantitative} or bacteria engulfed by phagocytes \cite{clarke2006phagocyte,kaufmann2016molecular}. Complex assemblies on a larger scale involve colloidal particles that are embedded within, e.g., droplets \cite{cho2005self,zwicker2017growth,zwicker2018positioning,grauer2021droploids} or vesicles \cite{paoluzzi2016shape,vutukuri2020active,peterson2021vesicle,willemsalvarez2022prep}, or dock at surfaces \cite{kunti2021rational,palacci2013photoactivated,rozynek2016patchy,schmidt2019light}.

Despite this plethora of realizations of particle uptake or encapsulation, the collective properties of larger assemblies of such interacting particle mixtures have not yet been systematically explored.
Even with simple pair interactions, the formation of carrier-cargo complexes has barely been considered from the angle of classical statistical mechanics.
This is most likely due to the intrinsic complexity of internal degrees of freedom, needed for a basic description of particles which are swallowed (or ejected again) and thus continuously change
their role from freely floating to loaded cargo.
Therefore there is a principal need for minimal models which provide insight into the composition and structure of such carrier-cargo mixtures.

Here, we develop a controlled setting, which allows us to classify the occupation statistics of different carrier-cargo mixtures and predict their structural properties within a first-principles framework of statistical mechanics.
First, we devise a basic model involving hollow carriers with excluded-volume interactions to exemplify that the emerging complexes of carriers occupied by cargo can be efficiently considered as individual species.
 This equilibrium picture allows us to relate the occupation probability (or engulfment strength) directly to the partition functions of confined cargo particles and we therefore speak of a Boltzmann occupation law.
Second, in view of the variety of the soft matter zoo or applications in biology, we interpret the individual engulfment strength as an effective quantity that should take into account processes at the carrier membrane.
Third, we argue that non-equilibrium uptake of multiple cargo would typically not follow a Boltzmann law. This could lead to intriguing collective effects,
as exemplified by postulating an occupation law which enables a Bose-Einstein condensation outside the quantum world \cite{evans1996bose,meng2021magnetic,tevrugt2023microscopic}.

 The paper is arranged as follows.
We first outline in Sec.~\ref{sec_theory} a general rigorous mapping to an effective system.
This mapping is then applied to hard particles in Sec.~\ref{sec_evaluation},
where we determine exact properties of one-dimensional carrier-cargo mixtures and show that our theoretical treatment leads to highly accurate predictions in higher dimensions.
Moreover, we elaborate in Sec.~\ref{sec_applications} on the role of attractive or soft interactions and propose empirical applications to non-equilibrium particle uptake.
We then conclude in Sec.~\ref{sec_conclusions}.

\begin{figure}[t]
\begin{center}
\includegraphics[width=1.0\linewidth]{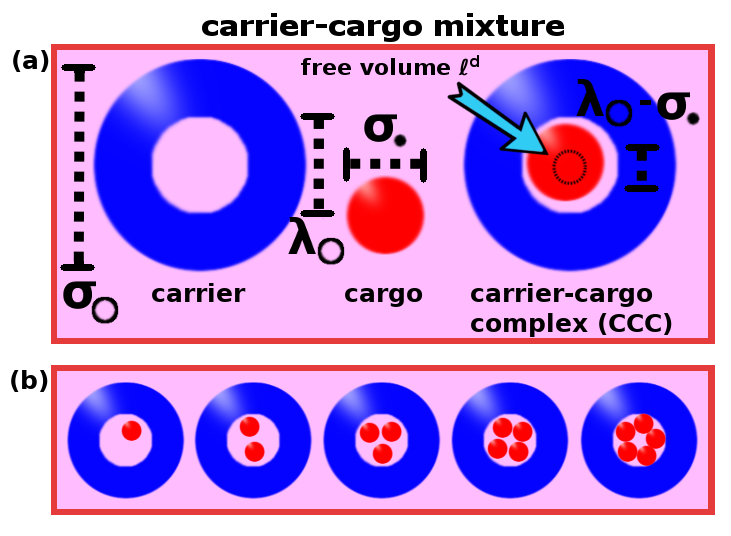}
\caption{\label{fig_concept_snap}
Hard-body model for complex-forming carrier-cargo mixtures.
(a) A hollow carrier (blue) and a smaller cargo (red) with annotated size parameters as basic building blocks.
The $d$-dimensional free volume $\ell^d$ (black circle) available to the center of the loaded cargo drives the formation of a carrier-cargo complex (CCC).
 For $\lambda_{\myCARRIER}-\sigma_{\myCARGO}<\sigma_{\myCARGO}$ there exists only one possible CCC.
(b) A carrier with larger interior ($\lambda_{\myCARRIER}-\sigma_{\myCARGO}>\sigma_{\myCARGO}$) can hold more than one cargo and thus different CCCs can form, illustrated here for occupation numbers $\nu=1,\ldots,5$.
}
\end{center}
\end{figure}

\section{Theoretical treatment \label{sec_theory}}

We consider two fundamental types of particles: carriers, which offer internal storage space, and cargo, which can occupy this space.
A carrier whose internal degrees of freedom are explored by cargo, represents a carrier-cargo complex (CCC).
For now, all particles are radially symmetric and interact solely via their excluded volume,
as described in Sec.~\ref{sec_model}.
After discussing in Sec.~\ref{sec_exact1d}, the exact solution for a one-dimensional excluded-volume model as a special case,
we motivate in Sec.~\ref{sec_combination} a general recipe for treating particle complexes irrespective of their interactions,
which we then apply in Sec.~\ref{sec_combiCC} to general carrier-cargo mixtures.
The notion of a CCC for more general interactions is discussed later in Sec.~\ref{sec_applications}.

\subsection{Excluded-volume model \label{sec_model}}

The ingredients of our minimal excluded-volume model in $d$ spatial dimensions are illustrated in Fig.~\ref{fig_concept_snap}.
We consider carriers with diameter $\sigma_{\myCARRIER}$, which possess a void space of diameter $\lambda_{\myCARRIER}$ in their interior,
and cargo with diameter $\sigma_{\myCARGO}$.
While particles of the same species $i\in\{\myCARRIER,\myCARGO\}$ interact with each other as $d$-dimensional hard spheres with the potential
\begin{equation}
\label{eq_Uijrod1}
    U_{ii} (r) =\left\{
\begin{array}{ll}
    0 & \textnormal{for } r \geq\sigma_i\,, \\
    \infty & \textnormal{else}\,,
\end{array}
\right.
\end{equation}
the interaction between a cargo and a carrier is given by
\begin{equation}
\label{eq_Uijrod2}
    U_{\myCARGO\myCARRIER} (r) =U_{\myCARRIER\myCARGO} (r) =\left\{
\begin{array}{ll}
    0 & \textnormal{for }r<  \left ( \lambda_\myCARRIER - \sigma_\myCARGO \right )/2\,, \\
    0 & \textnormal{for } r \geq\left ( \sigma_\myCARRIER + \sigma_\myCARGO \right )/2\,, \\
    \infty & \textnormal{else}\,,
\end{array}
\right.
\end{equation}
where $r$ is the center-to-center distance between a pair of particles.

\begin{figure}[t]
\begin{center}
\includegraphics[width=1.0\linewidth]{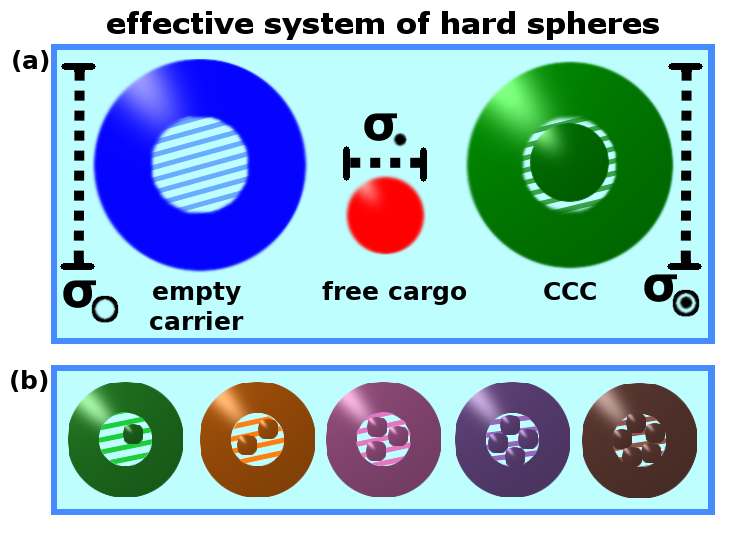}
\caption{\label{fig_concept_effective}
Effective description of the carrier-cargo mixtures depicted in Fig.~\ref{fig_concept_snap} as $d$-dimensional hard spheres.
Each CCC consisting of one carrier and exactly $\nu$ cargo particles is associated with an effective species (different colors).
All CCCs have the same physical properties, i.e., the same diameter $\sigma_\myCCC\equiv\sigma_\myCARRIER$, as the carriers (indicated by the shaded interior).
The internal degrees of freedom due to carrier occupation are mapped onto the effective chemical potential(s)
(a) $\mu_{\myCCC}$, defined in Eq.~\eqref{eq_3spec}, or
(b) $\mu_{\nu}$, defined in Eq.~\eqref{eq_geneffMUmu}, of the CCCs.
The empty carriers and free cargo have the chemical potentials $\mu_{\myCARRIER}$ and $\mu_{\myCARGO}$, respectively.}
\end{center}
\end{figure}

As illustrated in Fig.~\ref{fig_concept_snap}a, there exists a single possible CCC, representing a carrier holding exactly one cargo, if $\sigma_{\myCARGO}<\lambda_{\myCARRIER}<2\sigma_{\myCARGO}$.
For carriers with larger interiors, multiple CCCs can form, see Fig.~\ref{fig_concept_snap}b.
We will outline below, that the driving force of CCC formation is generally related to the standard Boltzmann statistics for a system of cargo particles in a cavity with the shape of the interior of the carrier.
Hence, the carrier occupation in our hard-body model is limited by close packing.
Our model is evaluated via Monte-Carlo simulation in the canonical ensemble with $M$ carriers and $N$ cargo particles, as described in Appendix~\ref{app_MC},
and, in the following, by statistical calculations in the grand canonical ensemble, where $\mu_{\myCARRIER}$ and $\mu_{\myCARGO}$ denote the chemical potentials of the two particle species.

\subsection{Exact solution in one dimension \label{sec_exact1d}}

To demonstrate step by step how to treat the general problem of CCC formation in an elegant way through effective chemical potentials,
we first consider the most intuitive model for a carrier-cargo mixture in which the carriers can hold at most one cargo.
We further focus for the moment on one spatial dimension where it is exactly solvable.
For this setup, we define the free length
\begin{equation}
\ell:=\lambda_{\myCARRIER}-\sigma_{\myCARGO}
\label{eq_epsilondef}
\end{equation}
of one loaded cargo within a carrier, such that $\ell<\sigma_{\myCARGO}$.
Note that also the case with $\lambda_{\myCARRIER}>2\sigma_{\myCARGO}$, allowing for multiple carrier occupation, is exactly solvable in one dimension, but for clarity of the following presentation, we will discuss this scenario later in Sec.~\ref{sec_combination}.

\subsubsection{Canonical partition function \label{sec_c3}}

As a first step we provide the exact canonical partition function $Z_{M,N}$ of the mixture consisting of $M$ carriers and $N$ cargo particles.
The standard partition function $\mathcal{Z}^{(L)}_{N_1,N_2,\ldots, N_K}(\sigma_1,\sigma_2,\ldots,\sigma_K)$ of a $K$-component hard rod mixture in a system of length $L$,
consisting of $N_i$ particles of length $\sigma_i$ for species $i=1,\ldots,K$, is stated in Appendix \ref{app_Z} as a reference.
Denoting by $C\leq\min(M,N)$ the number of CCCs (the carrier particles which are occupied by a cargo particle), we find
\begin{align}\label{eq_ZMN0}
Z_{M,N} &=\sum_{C=0}^{\min(M,N)} \mathcal{Z}^{(L)}_{M,N-C}(\sigma_{\myCARRIER},\sigma_{\myCARGO})\frac{\ell^C}{\Lambda^CC!}\frac{M!}{(M-C)!}\,,
\end{align}
where $\Lambda$ is the thermal wave length.
Each term in this sum corresponds to a number $N-C$ of free cargo particles interacting with the $M$ carriers as hard rods, while a multiplicative factor accounts for the occupation statistics of the $C$ bounded cargo particles within the CCCs.

Substituting $\mathcal{Z}^{(L)}_{M,N}$ into Eq.~\eqref{eq_ZMN0} we get
\begin{align}\label{eq_ZMN}
Z_{M,N}&=\sum_{C=0}^{\min(M,N)}\frac{\left(L-M\sigma_{\myCARRIER}-(N-C)\sigma_{\myCARGO}\right)^{(M+N-C)} \ell^C}{\Lambda^{M+N}(M-C)!(N-C)!C!}\,.
\end{align}
Although the sum over $C$ cannot be explicitly calculated, this result suggests an alternative interpretation of the binary carrier-cargo mixture as an effective three-component mixture, illustrated in Fig.~\ref{fig_concept_effective}.
As shown in Appendix~\ref{app_Z1dfact}, this effective mixture consists of $A:=M-C$ empty carriers, $B:=N-C$ free cargo particles and $C$ occupied carriers, i.e., CCCs.
This becomes intuitive when expressing Eq.~\eqref{eq_ZMN} solely in terms of the partition functions $\mathcal{Z}^{(L)}_{A,B,C}(\sigma_{\myCARRIER},\sigma_{\myCARGO},\sigma_{\myCARRIER})$ and $\mathcal{Z}^{(\lambda_{\myCARRIER})}_1(\sigma_{\myCARGO})=\ell/\Lambda$.

\subsubsection{Grand canonical partition function \label{sec_gc3}}

 The interpretation of the CCCs as members of a third particle species with $\ell$ being an intrinsic control parameter, becomes more transparent when switching to the grand-canonical picture.
 The exact grand canonical partition function
 \begin{align}\label{eq_Xi0}
\Xi&=\sum_{M=0}^\infty\sum_{N=0}^\infty\sum_{C=0}^{\min(M,N)} \frac{z_{\myCARRIER}^Mz_{\myCARGO}^N\ell^C}{(M-C)!(N-C)!C!}\\\nonumber
&\ \ \ \ \ \ \ \ \ \ \ \ \ \ \ \ \ \ \ \ \ \ \times \left(L-M\sigma_{\myCARRIER}-(N-C)\sigma_{\myCARGO}\right)^{(M+N-C)}\,
\end{align}
 of the carrier-cargo mixture can be determined from weighting Eq.~\eqref{eq_ZMN} with the fugacities $z_i:=e^{\beta\mu_i}
 /\Lambda$ of species $i\in\{\myCARRIER,\myCARGO\}$,
 where $\mu_i$ are the respective chemical potentials and $\beta=(k_\text{B}T)^{-1}$ is the inverse of the temperature $T$ with Boltzmann's constant $k_\text{B}$.
 Recognizing the identity
 \begin{align}
\sum_{M=0}^\infty\sum_{N=0}^\infty\sum_{C=0}^{\min(M,N)}
=\sum_{C=0}^\infty\sum_{M=C}^\infty\sum_{N=C}^\infty
\end{align}
 for the infinite series and shifting the indices through the substitutions $M\rightarrow A+C$ and $N\rightarrow B+C$,
 Eq.~\eqref{eq_Xi0} can be further evaluated as
\begin{align}\label{eq_Xi}
\Xi&=\sum_{A=0}^\infty\sum_{B=0}^\infty\sum_{C=0}^\infty \frac{z_{\myCARRIER}^Az_{\myCARGO}^B(z_{\myCARRIER}z_{\myCARGO}\ell)^C}{A!B!C!}\\\nonumber
&\ \ \ \ \ \ \ \ \ \ \ \ \ \ \ \ \ \ \times\left(L-A\sigma_{\myCARRIER}-B\sigma_{\myCARGO}-C\sigma_{\myCCC}\right)^{(A+B+C)}\,,
\end{align}
where we have introduced the diameter $\sigma_{\myCCC}\equiv\sigma_{\myCARRIER}$ of the CCCs for later convenience.
As in Sec.~\ref{sec_c3}, the values of $A$, $B$ and $C$ can be interpreted as the particle numbers associated with the different species.

\subsubsection{Exact mapping onto an effective ternary mixture \label{sec_exactmapping3}}

Defining the effective fugacity
\begin{equation}
    {z_{\myCCC}}:=\ell z_{\myCARRIER}z_{\myCARGO}
    \label{eq_3specFUG}
\end{equation}
associated with the third species of CCCs,
we have established a proper mapping of the two-component carrier-cargo mixture
onto a three-component mixture, illustrated in Fig.~\ref{fig_concept_effective}, of three (reacting) species:
empty carriers, controlled by the chemical potential $\mu_\myCARRIER$,
free cargo, controlled by the chemical potential $\mu_\myCARGO$,
and CCCs composed of one carrier and one cargo, with the effective chemical potential
\begin{align}
\mu_\myCCC:=k_\text{B}T\ln(\ell/\Lambda)+\mu_\myCARRIER+\mu_\myCARGO\,.
    \label{eq_3spec}
\end{align}
The interpretation of this mapping is, that the average number of CCCs is determined by both the two external particle reservoirs, represented by $\mu_\myCARRIER$ and $\mu_\myCARGO$ of the carrier-cargo mixture,
and the internal degrees of freedom, represented by the free length $\ell$ explored by the cargo upon occupying a carrier.

While the sums in Eq.~\eqref{eq_Xi} can be explicitly carried out, we refrain here from doing so.
Instead, we emphasize that, as soon as an effective fugacity~\eqref{eq_3specFUG} or chemical potential~\eqref{eq_3spec} is specified,
the properties of the mixture can be explicitly evaluated using the vast toolbox from liquid-state theory \cite{hansenmcdonald}.
This fundamental mapping holds even if the grand-canonical partition function cannot be exactly determined and will be generalized later to multiple cargo loading.
In fact, to (approximately) describe a general interacting system, it is sufficient to know the combinatorics of the CCC formation to specify $z_{\myCCC}$ and thus $\mu_\myCCC$, as explored in more detail in Appendix~\ref{app_ideal}.

In this manuscript, we make use of the framework of classical density functional theory (DFT) \cite{Evans1979} introduced in Appendix~\ref{app_DFT}
In the special case of one-dimensional hard rods, considered so far, we can thus obtain the exact statistics
by evaluating the Percus functional \cite{percus1976equilibrium,vanderlickpercus1989mixture} for a mixture of empty carriers of length $\sigma_\myCARRIER$, free cargo of length $\sigma_\myCARGO$ and CCCs of the same length $\sigma_{\myCCC}\equiv\sigma_{\myCARRIER}$ as the carriers.

\subsubsection{Number densities \label{sec_homdensities}}

Focusing on a spatially homogeneous system, we can gain general insight into the carrier-cargo mixture from Eq.~\eqref{eq_Xi}.
Introducing the homogeneous density operators $\hat{\rho}_X=\frac{X}{L}$, where $X\in\{M,N,A,B,C\}$ represents the number of particles in the different (effective) species,
the probabilities of aggregation of the carrier and cargo into a CCC can be determined from the ensemble average (grand-canonical trace) $\langle\hat{\rho}_X\rangle$ of $\hat{\rho}_X$.
Specifically, the total densities $\rhomyCARRIER:=\langle\hat{\rho}_M\rangle$ of all carriers and $\rhomyCARGO:=\langle\hat{\rho}_N\rangle$ of all cargo particles in the carrier-cargo mixture can be calculated from Eq.~\eqref{eq_Xi0}
and the densities $\rhoA:=\langle\hat{\rho}_A\rangle$ of empty carriers, $\rhoB:=\langle\hat{\rho}_B\rangle$ of free cargo and $\rho_\myCCC:=\langle\hat{\rho}_C\rangle$ of CCCs follow from Eq.~\eqref{eq_Xi}.

Performing the same manipulations which were used to derive Eq.~\eqref{eq_Xi} from Eq.~\eqref{eq_Xi0},
it is easy to show that
\begin{align}
\rhomyCARRIER=\rhoA+\rho_\myCCC\ \ \ \mbox{and}\ \ \ \rhomyCARGO=\rhoB+\rho_\myCCC
    \label{eq_densities}
\end{align}
These relations hold for any kind of interactions between the particles and also for a spatially inhomogeneous system.
It is thus generally possible to recover information on the physical system from an effective DFT calculation based on the mapping in Eq.~\eqref{eq_3spec}.

Moreover, if one is only interested in ratios of the (homogeneous) densities, it suffices to know the respective fugacities.
For example, the fraction $\rho_\myCCC/\rhomyCARRIER$ of CCCs, indicating the percentage of occupied carriers, follows as $\rho_\myCCC/\rhomyCARRIER=z_{\myCCC}/z_{\myCARRIER}^\text{t}$.
A concise discussion can be found in Appendix~\ref{app_ideal}.

\subsubsection{Pair distributions \label{sec_pairdistributions}}

To extract structural information on the carrier-cargo mixture from a standard calculation in the effective system,
the relation, Eq.~\eqref{eq_densities}, between the number densities can also be generalized to pair distribution functions,
 which indicate the probability to find two particles of a certain species at distance $x=|x_1-x_2|$.
This can be achieved by use of the additivity of the two-body densities while taking into account that a CCC represents both a carrier and a cargo with blurred position.

As further explained in Appendix~\ref{app_gtrue}, it is possible to express the pair distributions ${g}^\text{t}_{ij}(x)$ with $i,j\in\{\myCARRIER,\myCARGO\}$ of the physical carrier-cargo mixture
in terms of the (exactly known \cite{santos2007exact}) effective pair distributions ${g}_{ij}(x)$ with \mbox{$i,j\in\{\myCARRIER,\myCARGO,\myCCC\}$} of a three-component hard-rod mixture as
\begin{align}\label{eq_gactA}
g^\text{t}_{\myCARRIER\myCARRIER}&=g_{\myCARRIER\myCARRIER}\,,\\\label{eq_gactB}
g^\text{t}_{\myCARGO\myCARGO}&=\frac{\rhoB^2g_{\myCARGO\myCARGO}+\rhoB\rho_\myCCC\left(g^\text{(b)}_{\myCARGO\myCCC}+g^\text{(b)}_{\myCCC\myCARGO}\right)+\rho_\myCCC^2g^\text{(bb)}_{\myCCC\myCCC}}{(\rhomyCARGO)^2}\,,\\\label{eq_gactC}
g^\text{t}_{\myCARRIER\myCARGO}&=\frac{\rhoA\rhoB g_{\myCARRIER\myCARGO}+\rhoA\rho_\myCCC g^\text{(b)}_{\myCARRIER\myCCC}+\rhoB\rho_\myCCC g_{\myCCC\myCARGO}+G^\text{(b)}_{\myCCC\myCCC}}{\rhomyCARRIER\rhomyCARGO}\,,\\
g^\text{t}_{\myCARGO\myCARRIER}&=g^\text{t}_{\myCARRIER\myCARGO}\,.
\label{eq_gactD}
\end{align}
Here, the functions
\begin{align}\label{eq_gb}
g^\text{(b)}_{\myCCC l}
&:=\frac{1}{\ell}\int\mathrm{d}x' \Theta(\ell/2-|x_1-x'|) g_{\myCCC l}(|x'-x_2|)
=g^\text{(b)}_{l \myCCC}
\,,\\\label{eq_gbb}
g^\text{(bb)}_{\myCCC\myCCC}
&:=\frac{1}{\ell^2}\int\mathrm{d}x'\int\mathrm{d}x'' \Theta(\ell/2-|x_2-x''|)\\\nonumber & \ \ \ \ \ \ \ \ \ \ \ \ \ \ \ \ \ \ \ \times g_{\myCCC\myCCC}(|x'-x''|)
\Theta(\ell/2-|x_1-x'|) \,
\end{align}
with $l\in\{\myCARRIER,\myCARGO,\myCCC\}$ denote the blurred effective distributions and
\begin{align}\label{eq_Gbb}
\!\!\!\!\!\!G^\text{(b)}_{\myCCC\myCCC}
&:=\frac{\rho_\myCCC}{\ell} \Theta(\ell/2-|x_2-x_1|) +\rho_\myCCC^2
g^\text{(b)}_{\myCCC\myCCC}(|x_1-x_2|)\,\!\!\!
\end{align}
denotes the blurred effective two-body density of two CCCs extended by a blurred self contribution.

\subsection{Combinatorics of general particle complexes \label{sec_combination}}

As a next step, we extend our treatment~\eqref{eq_3spec} to general $d$-dimensional mixtures of $\kappa$ different particle species, representing different types of both carriers and cargo.
To this end, let us recall from Sec.~\ref{sec_exactmapping3} that it is sufficient to consider ideal point-like particles which only interact by forming complexes and establish the mapping for such a system.
The notion of carrier and cargo particles then follows by assigning appropriate interactions in the effective system.
In a more abstract manner, we can in general simply speak of free particles which can join to form a complex particle with internal degrees of freedom.
This allows for further applications, for example, in the context of aggregation or clustering.
In what follows, we continue using the term CCC when referring to any kind of effective complex particle.

\subsubsection{General properties of a CCC \label{sec_propertiesCCC}}

In a general mixture of $\kappa$ components, the total number $k$ of possible CCCs depends on both $\kappa$ and the particular occupation statistics of the carriers determined by the physical interactions in the system.
To establish the underlying combinatorics, let us  denote the particle number, chemical potential and fugacity of species \mbox{$i=1,\ldots,\kappa$} as $N_i$, $\mu_i$ and $z_i=e^{\beta\mu_i}/\Lambda^d$, respectively.
Now suppose a generic CCC of effective species $\nu=1,\ldots,k$ is made up from joining in total $X_\nu=\sum_i x_\nu^{(i)}$ building blocks,
where the $x_\nu^{(i)}$ denote the numbers of particles of species $i$ contributing to that particular type of CCC.

Such a CCC has $d$ external configurational degrees of freedom, associated with its spatial coordinates.
The remaining $(X_\nu-1)d$ internal configurational degrees of freedom act as a statistical weight for the aggregation and are specified by the underlying model.
For example, in our hard-body model for carrier-cargo mixtures, we discuss in Sec.~\ref{sec_genHB} that these correspond to the free volume available for the loaded cargo particles (or their canonical partition function for more general interactions).
Therefore, we choose for each $\nu$ an effective length scale $\ell_\nu$ counting the number of states associated with one internal degree of freedom.
 In other words, the occupation parameters $\ell_\nu$ follow from an occupation law of the carriers,
which for equilibrium carrier-cargo mixtures always obeys standard Boltzmann statistics.
More general occupation laws are suggested and discussed in Sec.~\ref{sec_occupationlaws}.

\subsubsection{Exact mapping onto a general effective mixture}

In generalization of Eq.~\eqref{eq_3specFUG}, the occurrence of general CCCs of species $\nu$ can be understood in terms of the effective fugacity
\begin{equation}
    z_\myCCC^{(\nu)}:=\ell_\nu^{(X_\nu-1)d} \prod_i^\kappa z_i^{x_\nu^{(i)}}\,,
    \label{eq_geneffFUG}
\end{equation}
as we show below in Sec.~\ref{sec_partitionGEN}.
Upon expressing $z_i$ and $z_\myCCC^{(\nu)}$ in terms of $\mu_i$ and $\mu_\myCCC^{(\nu)}=k_\text{B}T\ln\big(z_\myCCC^{(\nu)}\Lambda^d\big)$, respectively, we find the effective chemical potentials
\begin{align}
    \mu_\myCCC^{(\nu)}&:=k_\text{B}T(X_\nu-1)\,d\,\ln(\ell_\nu/\Lambda)+ \sum_{i=1}^\kappa x_\nu^{(i)}\mu_i\,.
    \label{eq_geneff}
\end{align}
This comprehensive combination law~\eqref{eq_geneff} can be utilized to describe any conceivable soft-matter system involving particle uptake or other types of bonding mechanisms
in terms of $\kappa+k$ effective components.

\subsubsection{General ideal partition functions \label{sec_partitionGEN}}

To derive Eq.~\eqref{eq_geneffFUG}, we consider point particles in a $d$-dimensional box of side length $L$.
For $\nu=1,\ldots,k$ let us denote by $C_\nu$ the number of CCCs of species $\nu$ that are present in a specific (allowed) configuration.
Then there are in total $C^{(i)}=\sum_{\nu=1}^k C_\nu x_\nu^{(i)}$ particles of species $i$ which contribute to any CCC, such that $N_i-C^{(i)}$ particles of that species remain free (or empty).
The condition that $N_i-C^{(i)}\geq0$ for all $i$ sets an upper bound to the numbers $C_\nu$ of CCCs in a given composition.

Now, taking the analogy to the considerations outlined in Sec.~\ref{sec_c3} and Appendix~\ref{app_Z1dfact},
we can write the canonical partition function as
\begin{align}\label{eq_ZGEN}
Z^\text{(id)}_{N_1,N_2,\ldots, N_\kappa}&=\sum_{\{C_\nu\}}^{\{N_i-C^{(i)}\geq0\}} \mathcal{Z}^{(L)}_{N_1-C^{(1)},\ldots,N_\kappa-C^{(\kappa)},C_1,\ldots,C_k}\!\!\!\!\!\!\!\!\!\!\!\!\!\!\! \cr
&\ \ \ \ \ \ \ \ \ \ \ \ \ \ \ \times \prod_{\nu=1}^k\left(\frac{\ell_\nu^{(X_\nu-1)}}{\Lambda}\right)^{dC_\nu}\,,\!\!\!\!\!\!\!\!\!
\end{align}
where the sum counts all sets of admissible numbers $\{C_\nu\}$ of CCCs such that the numbers $\{N_i-C^{(i)}\}$ of free particles are non-negative for all species.

Again, switching to the grand-canonical picture allows us to remove the interdependence of the sums. Subsequently introducing $A_i:=N_i-C^{(i)}$, we find the ideal
grand-canonical partition function
\begin{align}\label{eq_XiGEN}
\Xi^\text{(id)}&=\boldsymbol{\sum}_{AC}
\,\prod_{i=1}^\kappa \frac{(L^d\,z_i)^{A_i}}{A_i!}
\prod_{\nu=1}^k\frac{(L^d\,z_\myCCC^{(\nu)})^{C_\nu}}{C_\nu!}
\cr&=\exp\left(\sum_{i=1}^\kappa L^dz_i +\sum_{\nu=1}^k L^dz_\myCCC^{(\nu)}\right)\,.
\end{align}
with $\boldsymbol{\sum}_{AC}\equiv\sum_{A_1=0}^\infty\ldots\sum_{A_\kappa=0}^\infty\sum_{C_1=0}^\infty\ldots\sum_{C_k=0}^\infty$ and the effective fugacities $z_\myCCC^{(\nu)}$ from Eq.~\eqref{eq_geneffFUG}.
Such an effective $\kappa+k$-component mixture can then be equipped with physical interactions and evaluated accordingly.

\subsubsection{Number densities for general complexes}

As discussed in Sec.~\ref{sec_homdensities}, the composition of the effective mixture can be studied by
calculating the ensemble average of the density operators $\{\hat{\rho}_X\}$, where $X$ now represents $\{N_i\}$, $\{A_i\}$ and $\{C_\nu\}$.
It follows that the respective total number densities $\{\rho^\text{t}_{i}\}$ of particles of species $i$,
the number densities $\{\rho_i\}$ of free particles of species $i$ (which are not bound in a CCC) and
the number densities $\{\rho_\myCCC^{(\nu)}\}$ of CCCs of species $\nu$ are related by
\begin{equation}\label{eq_rhoMN_gen}
\rho^\text{t}_{i}=\rho_i+\sum_{\nu=1}^kx_\nu^{(i)}\rho_\myCCC^{(\nu)}\,,\ \ \ i\in\{1,\ldots,\kappa\}\,.
\end{equation}
In generalization of Eq.~\eqref{eq_densities}, the contribution of the CCCs to the total densities is weighted by the number $x_\nu^{(i)}$ of a single CCC's ingredients of species $i$.

\subsection{Combinatorics of carrier-cargo mixtures \label{sec_combiCC}}

As a special case of our general combinatorics established in Sec.~\ref{sec_combination}, we consider below a carrier-cargo mixture with $\kappa=2$ components, such that $i\in\{1,2\}$, which is equivalent to $i\in\{\myCARRIER,\myCARGO\}$ in our pictorial notation introduced in Sec.~\ref{sec_model}.
Specifically, a CCC of index $\nu=1,\ldots,k$ consists of $x_\nu^{(1)}=1$ carrier and $x_\nu^{(2)}=\nu$ cargo particles, which makes in total $X_\nu=1+\nu$ building blocks.
Below, we elaborate how our general results can be applied to this case.
Moreover, we discuss in Appendix~\ref{app_multiplecarriers} a more general mixtures involving $\kappa-1$ carrier species.

\subsubsection{Fugacities and number densities}

 Upon inserting $\kappa=2$, $x_\nu^{(1)}=1$ and $x_\nu^{(2)}=\nu$ into Eq.~\eqref{eq_geneffFUG} we get the effective fugacities
 \begin{equation}
    z_\myCCC^{(\nu)}
    =z_{\myCARRIER} (\ell_\nu^d z_{\myCARGO})^\nu\,,
    \label{eq_geneffMU}
\end{equation}
such that we further recover $z_\myCCC^{(1)}\simeq{z_{\myCCC}}$ from Eq.~\eqref{eq_3specFUG} in the special case $k=1$ with $\ell_1\simeq\ell$ for $d=1$.
Similarly, the effective chemical potentials from Eq.~\eqref{eq_geneff} become
\begin{align}
    \mu_\myCCC^{(\nu)}&=k_\text{B}T\nu\,d\,\ln(\ell_\nu/\Lambda)+ \mu_\myCARRIER+\nu\,\mu_\myCARGO
     \label{eq_geneffMUmu}
\end{align}
in accordance with Eq.~\eqref{eq_3spec}.

Accordingly, the relations in Eq.~\eqref{eq_rhoMN_gen} for the number densities become
\begin{equation}\label{eq_rhoMN_k}
\rhomyCARRIER=\rhoA+\sum_{\nu=1}^k\rho_\myCCC^{(\nu)}\,,\ \ \ \rhomyCARGO=\rhoB+\sum_{\nu=1}^k\nu\rho_\myCCC^{(\nu)}\,,
\end{equation}
where we again recover Eq.~\eqref{eq_densities} for $k=1$ with $\rho_\myCCC^{(1)}\simeq\rho_\myCCC$.
Depending on the system of interest, it may be quite cumbersome to determine all $\rho_\myCCC^{(\nu)}$ in Eq.~\eqref{eq_rhoMN_k}.
It may further prove insightful to explicitly evaluate the sums and we discuss below a convenient way to do so.

\subsubsection{Total number densities of CCCs and loaded cargo \label{sec_totaldensities}}

To keep the following discussion general, we formally take the limit $k\rightarrow\infty$,
noting that $\rho_\myCCC^{(\nu)}=0$ if there exists no CCC that contains $\nu$ cargo particles, i.e., if there is no free space, $\ell_\nu=0$, to place $\nu$ cargo particles inside a carrier.
We can thus define the total number density
\begin{equation}\label{eq_rhoMN_inf_CCCfrac}
\rho_\myCCC:=\sum_{\nu=1}^\infty\rho_\myCCC^{(\nu)}\,
\end{equation}
of all CCCs (irrespective of their species)
in a way that is consistent with the definition of $\rho_\myCCC$ in Sec.~\ref{sec_homdensities} for $k=1$.
Moreover, we define
\begin{equation}\label{eq_rhoMN_inf_LCfrac}
\rho_\circledast:=\sum_{\nu=1}^\infty\nu\rho_\myCCC^{(\nu)}
\end{equation}
as the total number density of loaded cargo, which scales with the average occupation number of a single CCC and is, in general, different from the total number density of CCCs, specifically $\rho_\circledast\geq\rho_\myCCC$.
To further evaluate these total number densities, we make use of their relations to the fugacities, as exemplified in Appendix~\ref{app_ideal}.

First, we define the appropriate effective fugacity
 \begin{equation}
    {z_{\myCCC}} :=\sum_{\nu=1}^\infty z_{\myCCC}^{(\nu)}= z_{\myCARRIER} \sum_{\nu=1}^\infty (\ell^d_\nu z_{\myCARGO})^\nu\,
    \label{eq_geneffMUsum}
\end{equation}
of a generalized CCC by adding up the individual contributions from Eq.~\eqref{eq_geneffMU}.
Hence, all CCCs with the same physical length $\sigma_{\myCARRIER}$ can be treated as a single species with number density $\rho_\myCCC$, i.e., we may merely distinguish between an empty and a nonempty carrier.

Second, we define $z_{\myCARRIER}^\text{t}=z_{\myCARRIER}+z_{\myCCC}$ and use the scaling relations
  \begin{align}\label{eq_scalingdensfugGEN}
\frac{\rhomyCARRIER}{z_{\myCARRIER}^\text{t}}=\frac{\rhoA}{z_{\myCARRIER}}=\frac{\rho_\myCCC^{(\nu)}}{z_{\myCCC}^{(\nu)}}=\frac{\rho_\myCCC}{z_{\myCCC}}\,
\end{align}
 to determine the generalized CCC fraction $\rho_\myCCC/\rhomyCARRIER$, as
 \begin{align}\label{eq_CCCfraction}
\frac{\rho_\myCCC}{\rhomyCARRIER}=\frac{z_{\myCCC}}{z_{\myCARRIER}^\text{t}}
=\frac{-1+\sum_{\nu=0}^\infty (\ell^d_\nu z_{\myCARGO})^\nu}{\sum_{\nu=0}^\infty (\ell^d_\nu z_{\myCARGO})^\nu}\,.
\end{align}
and the loaded-cargo fraction $\rho_\circledast/\rhomyCARRIER$, i.e., the number of loaded cargo per carrier, as
  \begin{align}\label{eq_loadedcargofraction}
\frac{\rho_\circledast}{\rhomyCARRIER}=\frac{\sum_{\nu=1}^\infty \nu \,z_{\myCCC}^{(\nu)}}{z_{\myCARRIER}^\text{t}}
=\frac{\sum_{\nu=0}^\infty \nu(\ell^d_\nu z_{\myCARGO})^\nu}{\sum_{\nu=0}^\infty (\ell^d_\nu z_{\myCARGO})^\nu}\,.
\end{align}
 We stress that these ratios are independent of the specific interparticle interactions.
What is left to be done is to specify all $\ell_\nu$ for $\nu=1,\ldots,k$
and then evaluate Eqs.~\eqref{eq_CCCfraction} and~\eqref{eq_loadedcargofraction}
as the fundamental characteristics of the carrier-cargo mixture.

\subsubsection{Boltzmann occupation law in equilibrium \label{sec_genHB}}

As exemplified in Sec.~\ref{sec_c3} and Appendix~\ref{app_Z1dfact} for one-dimensional excluded volume interactions allowing only for a single type of CCC,
the canonical partition function $Z_{M,N}$ of a general carrier-cargo mixture can be expressed as an appropriate combination of auxiliary partition functions.
In particular, upon properly deriving Eq.~\eqref{eq_ZGEN} with $\kappa=2$ and $X_\nu-1=\nu$, the terms
\begin{equation}
    \left(\frac{\ell_\nu}{\Lambda}\right)^{\!\nu d}=\mathcal{Z}^{(\mathcal{I}_\myCARRIER)}_\nu(\mathcal{B}_{\myCARGO})\,,\ \ \ \nu=1,2,\ldots
    \label{eq_epsilonentropicGEN_ALLALLALL}
\end{equation}
emerge when accounting for a configuration in which $\nu$ cargo particles with a certain shape (denoted by $\mathcal{B}_\myCARGO$) are confined to a carrier with a certain interior shape (denoted by $\mathcal{I}_\myCARRIER$).
The weight of all possible configurations is given by the canonical partition function $\mathcal{Z}^{(\mathcal{I}_\myCARRIER)}_\nu(\mathcal{B}_{\myCARGO})$.
Hence, through Eq.~\eqref{eq_epsilonentropicGEN_ALLALLALL}, all occupation parameters $\ell_\nu$ characterizing the formation of CCCs in our equilibrium model
have an explicit interpretation in terms of Boltzmann statistics and we say that the carrier occupation follows a Boltzmann law.
We substantiate this general result by providing three relevant examples.

For spherically symmetric hard bodies in $d$ dimensions with $\sigma_{\myCARGO}<\lambda_{\myCARRIER}<2\sigma_{\myCARGO}$, as illustrated in Fig.~\ref{fig_concept_snap}a,
 $\ell_1^d$ represents the spherical free volume with the diameter $\ell={\lambda}_\myCARRIER-\sigma_\myCARGO$ given in Eq.~\eqref{eq_epsilondef}, while all other occupation parameters are zero.
Hence we have
\begin{align}
   \ell_1 &= \frac{\ell\sqrt{\pi}}{2} \left(\Gamma\left(\frac{d}{2}+1\right)\right)^{\!-\frac{1}{d}} \,,\cr
   \ell_\nu&=0\,,\ \ \  \nu>1
   \label{eq_occupationk1}
\end{align}
with the gamma function $\Gamma$.
Together with Eq.~\eqref{eq_geneffMU}, we have fully recovered the result of Sec.~\ref{sec_exact1d} in the special case $d=1$, where we have $\ell_1=\ell$.

Next we consider in our excluded-volume model the case of carriers which can hold at most $k$ cargo particles, as illustrated in Fig.~\ref{fig_concept_snap}b.
Specifically for $d=1$ this corresponds to particles with $k\sigma_{\myCARGO}<\lambda_{\myCARRIER}<(k+1)\sigma_{\myCARGO}$, such that
\begin{align}
    \ell_\nu^{\nu}&=\frac{(\lambda_{\myCARRIER}-\nu\sigma_{\myCARGO})^\nu}{\nu!}=\frac{(\ell-(\nu-1)\sigma_{\myCARGO})^\nu}{\nu!}\,\ \ \ \nu\leq k\,,\cr
    \ell_\nu^{\nu}&=0\,\ \ \ \ \ \ \ \ \ \ \ \ \ \ \ \ \ \ \ \ \ \ \ \ \ \ \ \ \ \ \ \ \ \ \ \ \ \ \ \ \ \ \ \ \, \nu> k\,.
    \label{eq_epsilonentropicGEN}
\end{align}
Together with Eq.~\eqref{eq_geneffMU} and an appropriate treatment of interactions, these occupation parameters allow for an exact description of this one-dimensional carrier-cargo mixture,
where Eq.~\eqref{eq_occupationk1} is recovered for $k=1$.

We conclude by discussing the particularly simple case of non-interacting cargo with $\sigma_{\myCARGO}=0$, sticking to $d=1$ for simplicity.
Due to the absence of interactions in the occupied carrier, we formally have $k\rightarrow\infty$ with all occupation parameters
\begin{align}
\ell_\nu= \frac{\ell}{(\nu!)^{\frac{1}{\nu}}}
   \label{eq_epsilonentropicID}
\end{align}
depending on a single length scale $\ell =\ell_1=\lambda_{\myCARRIER}$.
In this case, the expressions from Eqs.~\eqref{eq_geneffMUsum},~\eqref{eq_CCCfraction} and~\eqref{eq_loadedcargofraction} can be considerably simplified to
\begin{align}
   {z_{\myCCC}} =z_{\myCARRIER} \left(e^{\ell z_{\myCARGO}}-1\right)\,,\ \ \ \frac{\rho_\myCCC}{\rhomyCARRIER} =1-e^{-\ell z_{\myCARGO}}\,,\ \ \
   \frac{\rho_\circledast}{\rhomyCARRIER}=\ell z_{\myCARGO}\,,
   \label{eq_fractionsBO}
\end{align}
where we have used
 $\sum_{\nu=0}^\infty (\ell z_{\myCARGO})^\nu/\nu!=e^{\ell z_{\myCARGO}}$
 and $\sum_{\nu=1}^\infty (\ell z_{\myCARGO})^\nu/(\nu-1)!=\ell z_{\myCARGO}e^{\ell z_{\myCARGO}}$.
We notice that the CCC fraction exponentially approaches the limiting value $\rho_\myCCC/\rhomyCARRIER\rightarrow1$, i.e., a system in which all carriers are occupied, when taking the limit $z_{\myCARGO}\rightarrow\infty$.

\section{Results for hard bodies \label{sec_evaluation}}

To illustrate the results of our theoretical treatment from Sec.~\ref{sec_theory}, we study the effective hard-body mixture within classical density functional theory (DFT), as described in Appendix~\ref{app_DFT}.
This amounts to solving a system of $\kappa+k$ coupled algebraic equations (one for each component of the effective system).
 While our treatment based on Fundamental Measure Theory (FMT) \cite{hansen2006WBII,roth2010fundamental,roth2012} is exact in one spatial dimension \cite{percus1976equilibrium,vanderlickpercus1989mixture},
there exists no exact theory for interacting systems in higher spatial dimensions.
Hence, we also compare our results to Monte-Carlo simulations as described in Appendix~\ref{app_MC}.

Specifically, we study the composition, given by the fraction $\rho_\myCCC/\rhomyCARRIER$ or $\rho_\myCCC^{(\nu)}/\rhomyCARRIER$ of all CCCs or CCCs of species $\nu$, respectively, among the carriers, as introduced in Secs.~\ref{sec_homdensities} and~\ref{sec_totaldensities}, and the structure, characterized by the pair distribution $g^\text{t}_{ij}$ in the physical carrier-cargo mixture, as discussed in Sec.~\ref{sec_pairdistributions}.

\subsection{Single-cargo uptake without interactions \label{sec_idealuptake}}

For a mixture of interacting particles, it is not possible to determine an explicit solution for the CCC fraction $\rho_\myCCC/\rhomyCARRIER$ as a function of the total number densities $\rhomyCARRIER$ of the carriers and $\rhomyCARGO$ of the cargo, even if a carrier can hold no more than a single cargo.
Thus, before discussing these results, we consider a simplified model of a point-like carrier-cargo mixture with $\sigma_\myCARRIER=\sigma_\myCARGO=\sigma_\myCCC=0$.
However, we still assume a positive interior length scale $\ell_1>0$ of the carrier, such that the only interaction is by cargo uptake, see Sec.~\ref{sec_attraction} or Sec.~\ref{sec_occupationlaws} for possible interpretations of such a scenario.

In this particular case, the CCC fraction follows in the desired closed form
\begin{align}
 \frac{\rho_\myCCC}{\rhomyCARRIER} = \frac{1+\mathcal{L}(1+\mathcal{R})-\sqrt{1+2\mathcal{L}(1+\mathcal{R})+\mathcal{L}^2(1-\mathcal{R})^2}}{2\mathcal{L}}
      \label{eq_exactfractionNI}
\end{align}
as the explicit solution of the three equations given by the two relations in Eq.~\eqref{eq_densities} and ${\rho_{\myCCC}}:=\ell_1^d \rho_{\myCARRIER}\rho_{\myCARGO}$, which follows from Eq.~\eqref{eq_geneffMU} with $\nu=1$ upon identifying the fugacities as number densities in the absence of effective interactions.
Here, the dimensionless parameters are the weighted interior volume $\mathcal{L}:=\ell_1^d\rhomyCARRIER$, where $\ell_1$ is specified in $d$ spatial dimensions by Eq.~\eqref{eq_occupationk1},
and the cargo-to-carrier ratio $\mathcal{R}:=\rhomyCARGO/\rhomyCARRIER$.

From Eq.~\eqref{eq_exactfractionNI} we can directly infer the basic behavior and qualitatively understand the composition of carrier-cargo mixtures.
Specifically, the CCC fraction is a positive and monotone increasing function of both parameters $\mathcal{L}$ and $\mathcal{R}$, which vanishes for either $\mathcal{L}=0$ or $\mathcal{R}=0$.
In the limit $\mathcal{L}\rightarrow\infty$ of an infinitely large available space inside the carrier, we obtain $\rho_\myCCC/\rhomyCARRIER\rightarrow \min(\mathcal{R},1)$, i.e., all cargo particles are engulfed by a carrier, while not all carriers are occupied for small cargo-to-carrier ratio $\mathcal{R}<1$.
In turn, for an infinite number of available cargo particles, $\mathcal{R}\rightarrow\infty$, all carriers are occupied, $\rho_\myCCC/\rhomyCARRIER\rightarrow 1$.
This nicely confirms the intuition that the number of CCCs in the system increases upon increasing either the free space in the interior of the carriers or the total number of cargo in the system.
However, the approximation in terms of non-interacting particles does not provide any insight into the structure of carrier-cargo mixtures.

\begin{figure}[t]
\begin{center}
\includegraphics[width=1.0\linewidth]{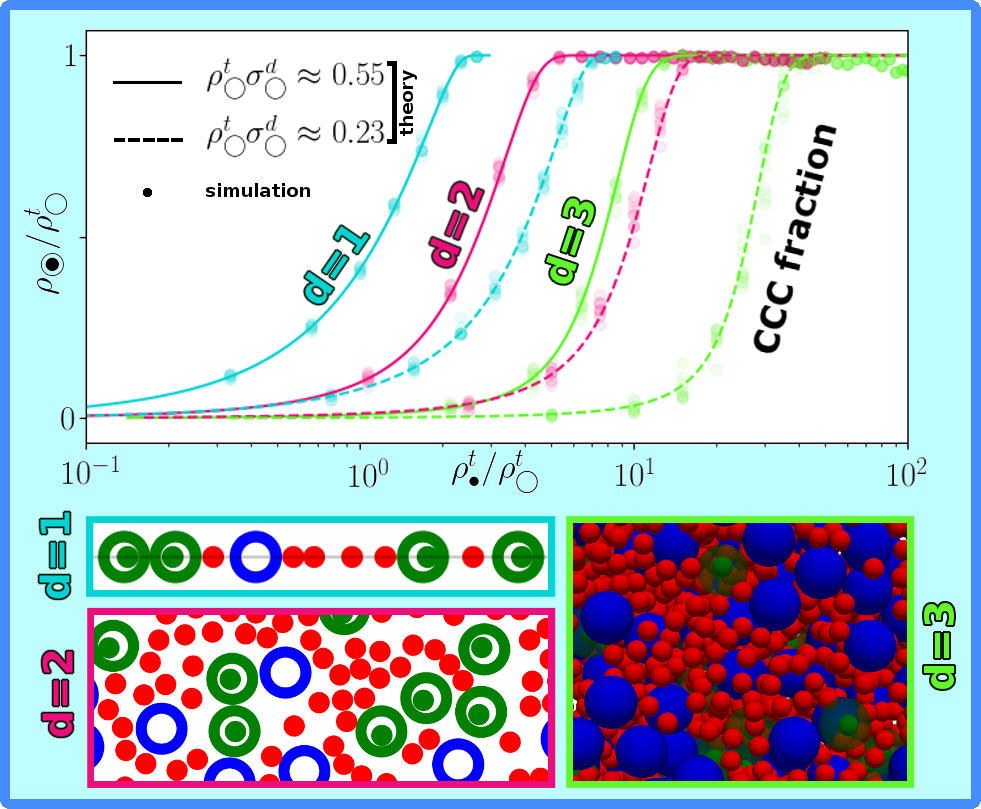}
\caption{\label{fig_CCCfractions123dp1}
Composition of a carrier-cargo mixture, cf.~Fig.~\ref{fig_concept_snap}a,
with $\sigma_\myCARGO=0.4\sigma_\myCARRIER$ and $\lambda_\myCARRIER=0.6\sigma_\myCARRIER$ determined from DFT (lines) and Monte-Carlo simulation (symbols).
We show the CCC fraction $\rho_\myCCC/\rhomyCARRIER$, indicating the percentage of occupied carriers, as a function of the total cargo density $\rhomyCARGO$ for different total carrier densities $\rhomyCARRIER$ (as labeled).
Our theory describes an effective system with the CCCs as a third species, compare Eq.~\eqref{eq_3spec}.
Below we show excerpts of typical simulation snapshots in which the CCCs are detected and colored in green, according to the scheme in Fig.~\ref{fig_concept_effective}a.
The one-dimensional system is illustrated as two-dimensional particles whose centers are confined to a line and the frames provide the color code for our results shown in $d=1,2,3$ dimensions.}
\end{center}
\end{figure}

\subsection{Single-cargo uptake in $d$ dimensions \label{sec_resultsk1}}

In our excluded-volume model, the behavior of a carrier-cargo mixture is governed by the competition between the overall gain of external free volume and the individual building blocks' sacrifice of entropy upon forming a CCC.
To gain further insight, we first discuss  the  statistical properties associated with carriers that can hold no more that a single cargo particle.
The CCC fraction $\rho_\myCCC/\rhomyCARRIER$ is shown in Fig.~\ref{fig_CCCfractions123dp1} as a function of the cargo-to-carrier ratio $\rhomyCARGO/\rhomyCARRIER$.
Also in spatial dimensions $d>1$, the results from our approximate DFT treatment relying on state-of-the art fundamental measure theory are in excellent agreement with the Monte-Carlo data.

Specifically, we predict in Fig.~\ref{fig_CCCfractions123dp1} that the fraction $\rho_\myCCC/\rhomyCARRIER$ of CCCs gradually increases with increasing cargo-to-carrier ratio $\rhomyCARGO/\rhomyCARRIER$ in any of the three considered spatial dimensions.
The same holds true when increasing the absolute number density $\rhomyCARRIER$ of all carriers while keeping $\rhomyCARGO/\rhomyCARRIER$ fixed, since an enhanced CCC formation balances the increase of global packing.
These results are qualitatively consistent with the analytical prediction in Eq.~\eqref{eq_exactfractionNI} for increasing $\mathcal{R}$ at constant $\mathcal{L}$ or vice versa.
Likewise, an increasing CCC fraction is observed for an increasing internal size $\ell=\lambda_\myCARRIER-\sigma_\myCARGO<$ of the carrier (not shown),
until multiple cargo uptake occurs for $\ell>\sigma_\myCARGO$ (see Sec.~\ref{sec_multipleuptakeres}).

Increasing the dimensionality of the system at fixed number densities,
the onset of the CCC formation in Fig.~\ref{fig_CCCfractions123dp1} shifts to higher cargo-to-carrier ratios.
This can be attributed on the one hand to the decreased packing efficiency of
hard-sphere systems for larger $d$, such that cargo can also occupy the voids between the carriers, thereby lowering the external drive towards CCC formation.
On the other hand, also the internal drive towards CCC formation is lowered for larger $d$, as the free volume $\ell_1^d$ of the carrier decreases relatively to $\sigma_\myCARRIER^d$, compare Eq.~\eqref{eq_occupationk1}.
Hence, also this dimensional aspect is qualitatively captured by Eq.~\eqref{eq_exactfractionNI}.

\begin{figure}[t]
\begin{center}
\includegraphics[width=1.0\linewidth]{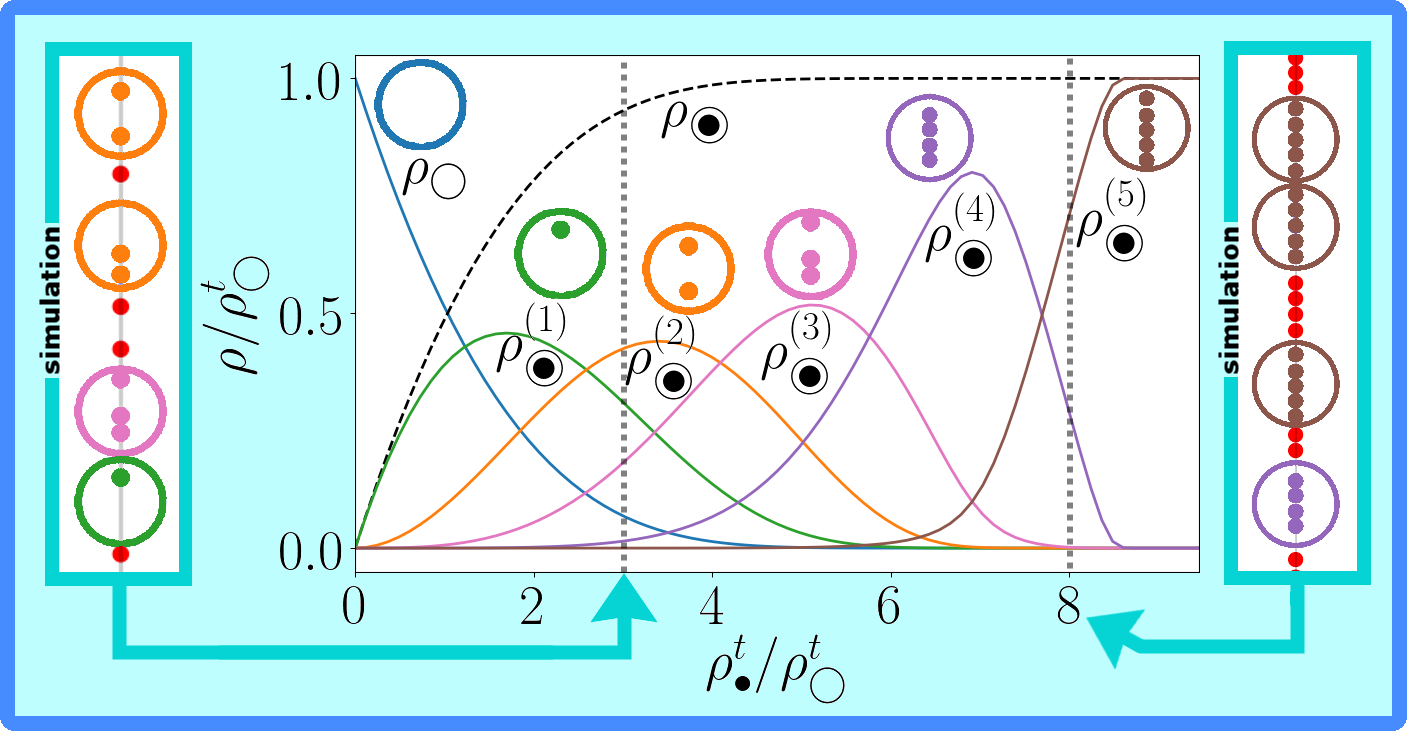}
\caption{\label{fig_multipleO}
Composition of a carrier-cargo mixture ($\kappa=2$) with multiple carrier occupation ($k=5$), cf.~Fig.~\ref{fig_concept_snap}b, for $d=1$.
 Different from Fig.~\ref{fig_CCCfractions123dp1}, we consider here $\rhomyCARRIER\sigma_\myCARRIER=0.25$ and the size parameters $\sigma_\myCARGO=\sigma_\myCARRIER/ 6$ and $\lambda_\myCARRIER=11\sigma_\myCARRIER/12$, such that up to $k=5$ cargo particles fit into one carrier.
The exact theoretical fraction of empty carriers, $\rho_\myCARRIER$, CCCs with $1\leq\nu\leq 5$ cargo, $\rho_\myCCC^{(\nu)}$, and total CCCs, $\rho_\myCCC=\sum_{\nu=1}^5\rho_\myCCC^{(\nu)}$, are shown as labeled, using the same coloring as in Fig.~\ref{fig_concept_effective}b.
The two boxes at the sides display typical excerpts from simulation snapshots, illustrated as in Fig.~\ref{fig_CCCfractions123dp1}, which reflect the plotted CCC fractions at the densities indicated by the teal arrows and dotted vertical lines.
}
\end{center}
\end{figure}

\subsection{Multiple-cargo uptake in one dimension \label{sec_multipleuptakeres}}

Turning to systems with carriers that offer more space to their cargo, we predict specifically for $d=1$ the exact density $\rho_\myCCC^{(\nu)}$ of each CCC representing $\nu=1,\ldots,k$ loaded cargo particles and the exact density $\rho_\myCCC$ of all CCCs according to Eq.~\eqref{eq_rhoMN_inf_CCCfrac}.
The corresponding CCC fractions are shown in Fig.~\ref{fig_multipleO} for a maximal carrier load $k=5$.
In this case, we see that the balance between internal and external entropical forces results in subsequent peaks of the different CCC fractions located at a higher total cargo density for larger $\nu$.
The simulation snapshots in Fig.~\ref{fig_multipleO} illustrate these percentages at the selected cargo-to-carrier ratios.
Due to the larger interior size of the carriers, compared to those in Sec.~\ref{sec_resultsk1} with $k=1$, the total CCC fraction $\rho_\myCCC/\rhomyCARRIER$ of non-empty carriers increases more rapidly than in Fig.~\ref{fig_CCCfractions123dp1} for a comparable density.

\begin{figure}[t]
\begin{center}
\includegraphics[width=1.0\linewidth]{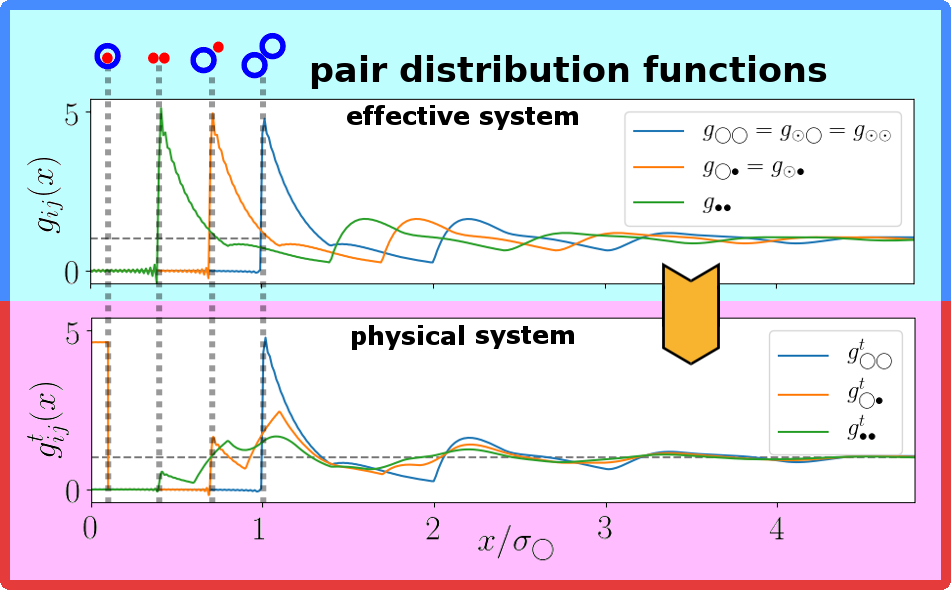}
\caption{\label{fig_CCCfractions123dp2}
Structure of the carrier-cargo mixture from Fig.~\ref{fig_CCCfractions123dp1}
with $\sigma_\myCARGO=0.4\sigma_\myCARRIER$ and $\lambda_\myCARRIER=0.6\sigma_\myCARRIER$ for $d=1$.
We show exact results obtained using DFT for the pair distribution functions at the densities $\rhomyCARGO\sigma_\myCARRIER=0.625$ and  $\rhomyCARRIER\sigma_\myCARRIER=0.72$.
The distributions for an effective hard-rod mixture (top) can be recombined to obtain the structure of the physical carrier-cargo mixture (bottom) according to Eqs.~\eqref{eq_gactA} to~\eqref{eq_gactD}.
By providing a direct measurement via Monte-Carlo simulation in Appendix~\ref{app_MC}, this result is exactly confirmed.
The lengths associated with the drawn characteristic two-particle configurations are highlighted by vertical lines.
}
\end{center}
\end{figure}

\subsection{Exact structure for one-dimensional single-cargo uptake \label{sec_resultsk1STRUCTURE}}

The possibility of spherical cargo to occupy the interior of hollow carriers is accompanied by structural changes of the carrier-cargo mixture compared to an ordinary hard-sphere mixture described by the effective system.
We illustrate the deeper connection between these two systems by discussing the exact  pair distributions for the one-dimensional particles considered in Sec.~\ref{sec_resultsk1}.
This can be conveniently achieved  by reconstructing the distributions ${g}^\text{t}_{ij}(x)$ with $i,j\in\{\myCARRIER,\myCARGO\}$ of the physical carrier-cargo mixture from the effective DFT results
${g}_{ij}(x)$ with $i,j\in\{\myCARRIER,\myCARGO,\myCCC\}$  according to Eqs.~\eqref{eq_gactA} to~\eqref{eq_gactD}.

While Fig.~\ref{fig_CCCfractions123dp2} illustrates the basic relation
$g^\text{t}_{\myCARRIER\myCARRIER}=g_{\myCARRIER\myCARRIER}$,
it also shows that $g^\text{t}_{\myCARRIER\myCARGO}$
possesses as a remarkable signature of CCC formation a strong signal for short distances.
Moreover, $g^\text{t}_{\myCARRIER\myCARGO}$ exhibits two (additional) and $g^\text{t}_{\myCARGO\myCARGO}$ even three main peaks, associated with those of a binary hard-sphere mixture.
The broadening of the peaks related to cargo within CCCs is apparent, as most clearly visible for $g^\text{t}_{\myCARGO\myCARGO}$ at the distance corresponding to two CCCs at contact.

\section{Applications beyond hard bodies \label{sec_applications}}

In Sec.~\ref{sec_evaluation} we have presented the main observations, derived in Sec.~\ref{sec_theory}, for our binary carrier-cargo mixture with excluded volume interactions in equilibrium.
 Our central theoretical result, the combination law in Eq.~\eqref{eq_geneffFUG} or Eq.~\eqref{eq_geneff}, is however, more versatile
as it offers the flexibility for investigating more complex interactions between the occurring (effective) species.
Moreover, we can formally assign an individual interaction to each CCC species which may be distinct from that of the carrier species
and interpret the $\ell_\nu$ as generalized heuristic parameters for controlling the formation of particle complexes.
In Sec.~\ref{sec_combinatoricsGENERAL}, we discuss more general carrier-cargo mixtures in the light of these points
 and identify in Sec.~\ref{sec_BEcond} a Bose-Einstein condensation from a speculative treatment of non-equilibrium particle uptake.
Applications beyond carrier-cargo mixtures are mentioned in our conclusions, Sec.~\ref{sec_conclusions}.

\subsection{Combinatorics of CCC formation \label{sec_combinatoricsGENERAL}}

 The combination law in Eq.~\eqref{eq_geneffMU} or Eq.~\eqref{eq_geneffMUmu} provides a general relation between the internal degrees of freedom of carrier-cargo mixtures
and an effective external particle reservoir.
It is exact for all equilibrium systems with interactions which clearly distinguish between the interior and the exterior of the carrier and thus allow for an unambiguous definition of CCCs.
Then, all $\ell_\nu$ carry all microscopic information on the physical driving forces behind particle uptake.

More generally, the driving force of a carrier to engulf its cargo may have different physical, chemical or biological origins, e.g., resulting from entropic, energetic, active or even intelligent uptake mechanisms.
Following heuristic intuition, a basic model for CCC formation should account for two internal properties:
(i) an engulfment strength, which quantifies the individual uptake probability and
(ii) a corresponding occupation law governing multiple cargo uptake.
 In our model, both properties can be accounted for by the parameters $\ell_\nu$, either rigorously, empirically or heuristically, as we discuss below for different physical scenarios.

\subsubsection{Interpretation of the excluded-volume model
\label{sec_interpretation}}

For completeness, we briefly recapitulate the situation for pure excluded-volume interactions.
In this case, the engulfment strength $\ell_1$, associated with the first cargo loaded, is given explicitly by its available $d$-dimensional free volume, compare Eq.~\eqref{eq_occupationk1}.
The engulfment strength $\ell_\nu$, associated with the $\nu$-th cargo loaded, is generally reduced (or even zero), since the occupation law is based on Boltzmann statistics, compare Eq.~\eqref{eq_epsilonentropicGEN}.
Even for non-interacting particles, there is still an effective repulsion of the cargo within the carrier upon multiple uptake due to the factor $1/(\nu!)$ in Eq.~\eqref{eq_epsilonentropicID}.
As a consequence of this indistinguishability, the tendency towards cargo uptake depends on the number of cargo already engulfed, which is a signature of the equilibrium nature of our model.

\subsubsection{Attraction-enhanced engulfment strength \label{sec_attraction}}

While hitherto we have exclusively used typical lengths $\ell_\nu$ to quantify cargo uptake,
we stress that our model is not limited to excluded-volume interactions.
As demonstrated by Eq.~\eqref{eq_epsilonentropicGEN_ALLALLALL}, the engulfment strength is rather associated with dimensionless partition functions of the confined cargo.
Thus an equally valid interpretation of engulfment strength is possible in terms of the binding energies
\begin{equation}
E_\nu:=-k_\text{B}Td\ln(\ell_\nu/\Lambda)\,.
\end{equation}
In general, we can describe cargo uptake, that is governed by a combination of soft interactions and available space inside vesicles, while both driving forces can then be commonly quantified by either $\ell_\nu$ or $E_\nu$.

Since attraction mechanisms are expected to play an important role in the process of cargo uptake,
we propose the modified carrier-cargo interaction potential
\begin{equation}
\label{eq_Uijrod2XNa}
  \!\!\!  U_{\myCARGO\myCARRIER} (r) =U_{\myCARRIER\myCARGO} (r) =\left\{
\begin{array}{ll}
    -a & \textnormal{for }r<  \left ( \lambda_\myCARRIER - \sigma_\myCARGO \right )/2\,, \\
    0 & \textnormal{for } r \geq\left ( \sigma_\myCARRIER + \sigma_\myCARGO \right )/2\,, \\
    \infty & \textnormal{else}\,
\end{array}
\right.\!\!\!\!
\end{equation}
as a basic example,
which extends Eq.~\eqref{eq_Uijrod2} by introducing the attraction parameter $a\geq0$.
 Apparently, upon evaluating Eq.~\eqref{eq_epsilonentropicGEN_ALLALLALL}, there is an additional Boltzmann factor $\exp(\beta a/d)$, which enhances the engulfment strengths $\ell_\nu$ according to setting $\ell_\nu\rightarrow\ell_\nu\exp(\beta a/d)$.
Therefore, a carrier-cargo mixture interacting according to Eq.~\eqref{eq_Uijrod2XNa} remains exactly solvable in one spatial dimension.
Strictly speaking, the idealized uptake scenario without interactions, discussed in Sec.~\ref{sec_idealuptake}, should be interpreted as a special limit of the situation described here.

This generalized consideration shows that the use of the parameters $\ell_\nu$ (which have the dimensions of length) is not limited to describe excluded-volume interactions, but allows for a quite general description of engulfment strength.
In particular, the value of $\ell_\nu$ can exceed the dimensions of the carrier.
Also with this interpretation, the parameters $\ell_\nu$ rigorously derive from a microscopic model, as specified here through Eq.~\eqref{eq_Uijrod2XNa},
which imposes a Boltzmann occupation law.

\subsubsection{Soft interactions and membranes \label{sec_soft}}

In many practical applications, cargo uptake is governed by soft interactions with a carrier membrane, which controls the encapsulation (and release) of particles.
Such a scenario without hard particle boundaries can still be described within our approach, albeit only in an approximate way.
To this end, let us denote by $\sigma_\myCARRIER/2$ the radius of the carrier membrane, which should be chosen to represent a (sufficiently high) potential maximum.
Extremely soft potentials which allow for a strong overlap of the carriers would be inappropriate for the purpose of identifying CCCs in the first place.
It is now convenient to split the pair interaction
\begin{equation}
\label{eq_Uijrod2XNb}
    U_{\myCARGO\myCARRIER} (r) =U_{\myCARRIER\myCARGO} (r) =\left\{
\begin{array}{ll}
    U_\text{in}(r) & \textnormal{for } r <\sigma_\myCARRIER/2\,, \\
    U_\text{out}(r) & \textnormal{else}\,,
\end{array}
\right.
\end{equation}
between a carrier and a cargo into the contributions $U_\text{in}(r)$ and $U_\text{out}(r)$ acting inside and outside the membrane, respectively.

Following Eq.~\eqref{eq_epsilonentropicGEN_ALLALLALL}, the engulfment strengths $\ell_\nu$ can be determined from the partition functions of the interacting cargo, with $U_\text{in}(r)$ acting as an external potential for $r<\sigma_\myCARRIER/2$.
The remaining task is to specify the interactions in the effective system.
To prevent the formal uptake of additional cargo (which is inconsistent with our effective picture of free cargo and empty carriers),
we must assume a hard core for $r <\sigma_\myCARRIER/2$ of both the empty carriers and the CCCs.
A first possibility to choose the interactions in the outside region is to simply take those with a carrier.
By doing so, we neglect direct interactions through the carrier membrane, which might be justified if all potentials are sufficiently short ranged.
Improved effective long-ranged interactions with a CCC can be specified by adding blurred versions of the interaction potentials with the engulfed cargo,
similar to what is done for the pair distributions in Eqs.~\eqref{eq_gb} to \eqref{eq_Gbb}.
Despite such approximations, the resulting effective treatment is still related to a microscopic model, as specified here through Eq.~\eqref{eq_Uijrod2XNb},
and, in particular, the parameters $\ell_\nu$ still obey a Boltzmann occupation law.

\subsubsection{Heuristic engulfment strength}

 In biological carrier-cargo systems, the complexity of physical processes that happen at the carrier membrane goes beyond refined equilibrium models based on soft potentials.
For example, a proper description should take into account adhesive forces, bending rigidity, surface tension or signaling.
Moreover, there exist mechanisms, such as the digestion of encapsulated cargo, which prevent its release.
This can apparently break detailed balance, an important underlying principle of equilibrium physics and Boltzmann statistics.

 Microscopic models, which accurately reflect all (non-equilibrium) driving forces for cargo uptake in such processes are presumably difficult to handle.
  To this end, we suggest to use our combinatorics as part of an effective treatment, in which the engulfment strengths $\ell_\nu$ represent free parameters.
While this strategy is apparently detached from the full microscopic information,
the freedom of choosing the $\ell_\nu$ underlines that the fundamental Boltzmann law of particle uptake may be broken for general carrier-cargo mixtures.
In other words, systems in nature suggest occupation laws that are not of the Boltzmann type.

\subsubsection{Alternative occupation laws from quantum statistics \label{sec_occupationlaws}}

\begin{figure}[t]
\begin{center}
\includegraphics[width=1.0\linewidth]{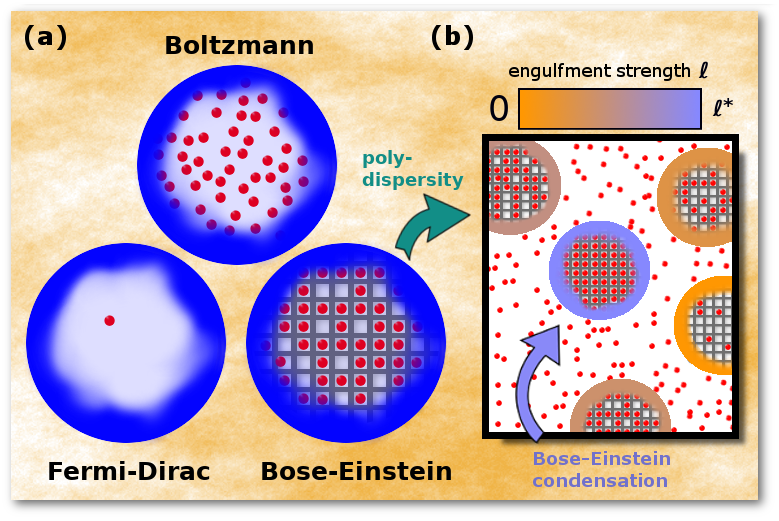}
\caption{\label{fig_concept_snap_occupation}
Heuristic occupation laws, suggesting the application of our methods for non-equilibrium particle uptake.
(a) The Boltzmann occupation law (top), a fundamental principle for equilibrium models, may be effectively broken in non-equilibrium systems.
Taking cues from quantum statistics, we postulate two alternatives.
 \textit{Fermi-Dirac occupation} (left) occurs if
a carrier is designed to engulf no more than a single cargo.
\textit{Bose-Einstein occupation} (right) formally corresponds to distinguishable cargo storage, illustrated here by drawing multiple compartments, each characterized by the same engulfment strength $\ell$.
(b) Polydispersity of the engulfment strength $\ell$ (indicated by the particle colors) can result in a Bose-Einstein condensation of cargo on the carriers with the maximal value $\ell=\ell^*$ (blue) if Bose-Einstein occupation is assumed.
}
\end{center}
\end{figure}

While it is challenging to systematically apply our methods to non-equilibrium particle uptake, we now aim to understand
the basic implications of effective non-Boltzmann occupation laws (arising from heuristically choosing the engulfment strengths) on multiple cargo uptake.
To this end, we assume that typical cargo can be much smaller than the carriers and consider again an idealized system with non-interacting cargo in $d=1$ dimensions.
Recalling the results in Eq.~\eqref{eq_fractionsBO} obtained with the ideal Boltzmann occupation law from Eq.~\eqref{eq_epsilonentropicID},
we consider two academic examples of alternative occupation laws inspired by quantum statistics, as illustrated in Fig.~\ref{fig_concept_snap_occupation}a.

A carrier that closes down after taking up a single cargo particle constitutes the soft-matter analogy to Pauli exclusion.
This Fermi-Dirac occupation law is defined by $\ell_1=\ell$ and $\ell_\nu=0$ for $\nu>1$, as in Eq.~\eqref{eq_occupationk1} but with $\ell$ being completely unrelated to any particle dimension or other interactions.
In this case, the expressions from Eqs.~\eqref{eq_geneffMUsum},~\eqref{eq_CCCfraction} and~\eqref{eq_loadedcargofraction} can simply be calculated by truncating all sums after $\nu=1$, which yields
\begin{equation}
    {z_{\myCCC}} = z_{\myCARRIER}\ell z_{\myCARGO}\,,\ \ \ \frac{\rho_\myCCC}{\rhomyCARRIER} =\frac{\ell z_{\myCARGO}}{1+\ell z_{\myCARGO}}\,,\ \ \ \frac{\rho_\circledast}{\rhomyCARRIER}=\frac{\ell z_{\myCARGO}}{1+\ell z_{\myCARGO}}\,.
    \label{eq_fractionsFD}
\end{equation}
In contrast to Boltzmann occupation, the limit $\rho_\myCCC/\rhomyCARRIER\rightarrow1$ for $z_{\myCARGO}\rightarrow\infty$ is approached algebraically as a function of $z_{\myCARGO}$.

If a carrier can store all cargo particles independently, or, more generally, if the engulfment strength does not depend on the number of cargo already engulfed by the carrier,
there is effectively no repulsion at all between the engulfed non-interacting cargo particles, as if each of them is stored in an individual compartment of size $\ell$.
Formally, this reflects the behavior of distinguishable particles (for example due to instant digestion),
 governed by a Bose-Einstein occupation law which is defined as
\begin{align}
\ell_\nu= \ell\,,\ \ \ \nu=1,2,\ldots
\label{eq_BElaw}
\end{align}
and which yields
\begin{equation}
   {z_{\myCCC}} =z_{\myCARRIER} \frac{\ell z_{\myCARGO}}{1-\ell z_{\myCARGO}}\,,\ \ \ \frac{\rho_\myCCC}{\rhomyCARRIER} = \ell z_{\myCARGO}\,,\ \ \
   \frac{\rho_\circledast}{\rhomyCARRIER}=\frac{\ell z_{\myCARGO}}{1-\ell z_{\myCARGO}}\,
   \label{eq_fractionsBE}
\end{equation}
with the restriction to $ \ell z_{\myCARGO}<1$.
Here we have used the identities
  $\sum_{\nu=0}^\infty (\ell z_{\myCARGO})^\nu=\frac{1}{1-\ell z_{\myCARGO}}$ and
 $\sum_{\nu=1}^\infty \nu(\ell z_{\myCARGO})^\nu=\frac{\ell z_{\myCARGO}}{(1-\ell z_{\myCARGO})^2}$
for the case $\ell z_{\myCARGO}<1$ and silently omitted including the step function $\Theta(1-\ell z_{\myCARGO})$ in these formulas.
For larger $\ell z_{\myCARGO}>1$ the above quantities are ill-defined, as both the effective fugacity and the loaded-cargo fraction diverge for $\ell z_{\myCARGO}\rightarrow1$.
The CCC fraction approaches the limiting value $\rho_\myCCC/\rhomyCARRIER\rightarrow1$ already at the finite value $z_{\myCARGO}\rightarrow\ell^{-1}$ of the fugacity of the cargo particles.

Comparing the different scenarios \eqref{eq_fractionsBO} \eqref{eq_fractionsFD} \eqref{eq_fractionsBE}, it can be shown that Fermi-Dirac (Bose-Einstein) occupation yields the smallest (largest) values of both $\rho_\myCCC$ and $\rho_\circledast$ and the weakest (strongest) increase of the CCC fraction, $\rho_\myCCC/\rhomyCARRIER$ as a function of $\rhomyCARGO$.
One intriguing result is that, in each case, the number of loaded cargo per carrier, $\rho_\circledast/\rhomyCARRIER$,
takes the same functional form as the average occupation number of a certain energy level in quantum statistics.

\begin{figure}[t]
\begin{center}
\includegraphics[width=0.97\linewidth]{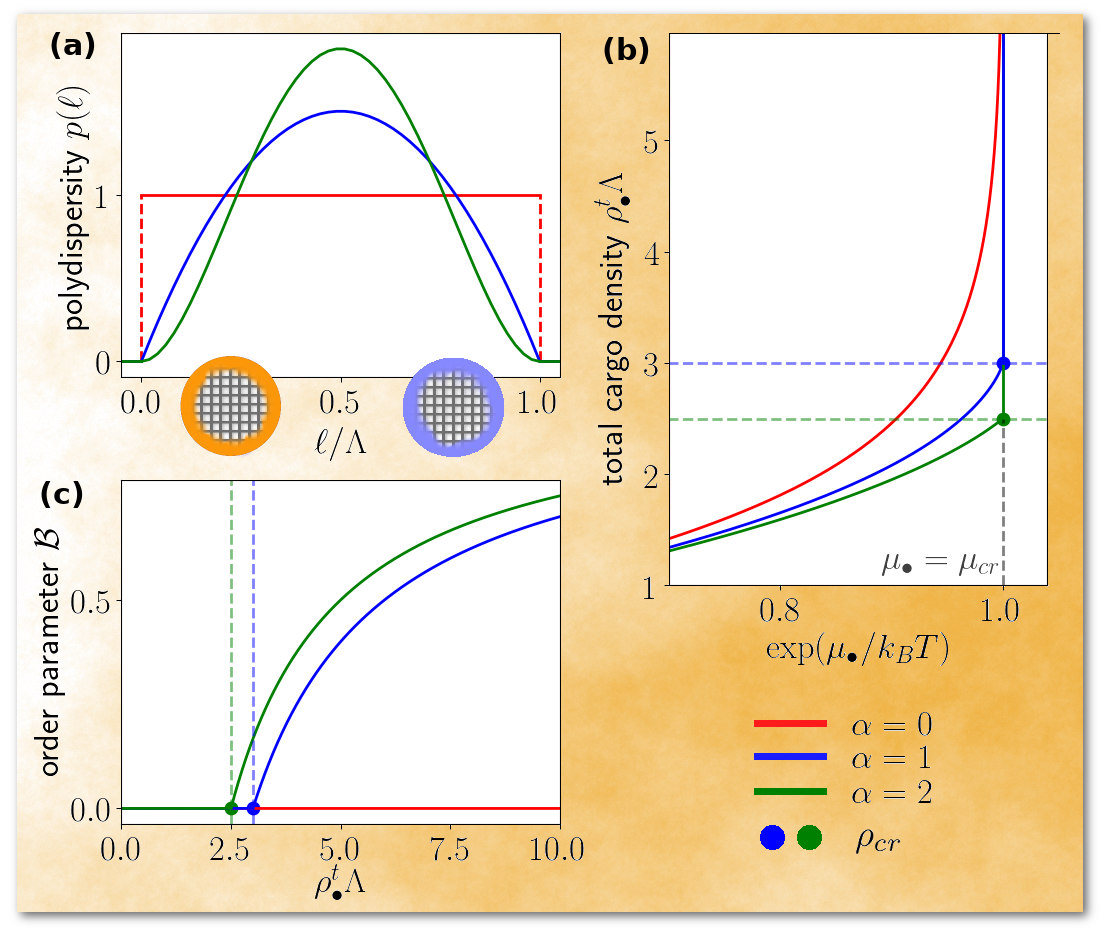}
\caption{\label{fig_BEC}
Bose-Einstein condensation for carriers with a supposed Bose-Einstein occupation law and polydisperse engulfment strength $\ell$ with $0\leq \ell\leq\ell^*:=\Lambda$.
We assume a polydispersity given by the function $p(\ell)\propto \ell^\alpha(\ell^*-\ell)^\alpha$
with representative exponents $\alpha=0,1,2$ (colors as labeled) and the total carrier density is $\bar{\rho}_\myCARRIER^\text{t}\Lambda=1$. (a) Normalized distributions $p(\ell)$ and illustrations of
two empty carriers with different engulfment strengths according to the color scheme in~Fig.~\ref{fig_concept_snap_occupation}b.
(b) The total cargo density $\rhomyCARGO$ increases with the chemical potential $\mu_\myCARGO$.
At $\mu_\myCARGO=\mu_\text{cr}$ (vertical line) $\rhomyCARGO$ diverges if $\alpha\leq1$ and reaches the finite critical value $\rho_\text{cr}$ (bullets) otherwise. (c) Order parameter~$\mathcal{B}$~(Eq. \eqref{eq_rhoXBE}) of the Bose-Einstein condensation, i.e., fraction of cargo on carriers with largest engulfment strength $\ell^*$, as a function of $\rhomyCARGO$.}
\end{center}
\end{figure}

\subsection{Bose-Einstein condensation with polydispersity \label{sec_BEcond}}

To demonstrate the collective effects that can, in principle, arise from an effective treatment of non-equilibrium cargo uptake, let us elaborate on a particular example.
Suppose that a Bose-Einstein occupation law according to Eq.~\eqref{eq_BElaw} does apply,
we additionally consider a polydisperse mixture of carriers, each with a characteristic engulfment strength $0\leq\ell\leq\ell^*$.
Then, as illustrated in Fig.~\ref{fig_concept_snap_occupation}b,
a Bose-Einstein condensation of the cargo can occur on the carriers with the strongest drive to engulf their cargo, represented by $\ell=\ell^*$.
The conditions for this phase transition are elaborated below and exemplified in Fig.~\ref{fig_BEC}.

The following considerations are independent of the interactions between the particles.
However, our working hypothesis assumes that cargo particles become distinguishable and non-interacting upon being engulfed, e.g., as an effective picture for being gradually digested.

\subsubsection{Polydisperse carriers}

To consider a  mixture of different carrier species $i$, the definitions in Sec.~\ref{sec_totaldensities} can be easily generalized, as performed in Appendix~\ref{app_multiplecarriers}.
In particular, the engulfment strengths $\ell^{[i]}$, and thus the resulting densities of (empty) carriers, CCCs and loaded cargo now additionally depend on the carrier species $i$.
Here, we are specifically interested in the total number densities $\rho_\circledast^{[i]}$ of cargo loaded on carriers of species $i$ under Bose-Einstein occupation,
which we can express from the result in Eq.~\eqref{eq_fractionsBE} as
\begin{align}\label{eq_BEfracI}
 \rho_\circledast^{[i]}=\frac{\ell^{[i]} z_{\myCARGO}}{1-\ell^{[i]} z_{\myCARGO}}\,\rho_{\myCARRIER}^{([i],\text{t})}
\end{align}
with the restriction to $\ell^{[i]} z_{\myCARGO}<1$.

For a mixture of carriers with polydisperse engulfment strengths $\ell$ the quantities of interest become continuous functions of $\ell$.
As input, we suppose that the total number density $\rhomyCARRIER(\ell)=\bar{\rho}_\myCARRIER^\text{t} p(\ell)$ of carriers is distributed according to a known normalized function $p(\ell)$, which vanishes beyond a maximal size $\ell^*$, see Fig.~\ref{fig_BEC}a for three examples considered here.
The total number density
$\bar{\rho}_\myCARRIER^\text{t}=\int_0^{\ell^*}\mathrm{d}\ell\,\rhomyCARRIER(\ell)/\ell^*$
of all carriers follows by integration (and we write here $\bar{\rho}_\myCARRIER^\text{t}$ instead of $\rho_\myCARRIER^\text{t}$ to indicate the averaging involved).
With the knowledge of $p(\ell)$, we can take the continuum limit of Eq.~\eqref{eq_BEfracI} to express the total number density $\rho_\circledast(\ell)$ of cargo loaded on carriers of engulfment strength $\ell$ under Bose-Einstein occupation as
\begin{align}\label{eq_BEfracL}
 \rho_\circledast(\ell)=\bar{\rho}_{\myCARRIER}^{\text{t}}\,p(\ell)\,\frac{\ell z_{\myCARGO}}{1-\ell z_{\myCARGO}}
\end{align}
with the restriction $\ell z_{\myCARGO}<1$.

Then, the total number of carriers is given by
\begin{equation} \label{eq_rhotBE}
\rhomyCARGO =\rhoB +\frac{\bar{\rho}_{\myCARRIER}^{\text{t}}}{\ell^*}\int_0^{\ell^*}\mathrm{d}\ell\,p(\ell)\,
\frac{\ell z_{\myCARGO}}{1-\ell z_{\myCARGO}}\,\Theta(1-\ell z_{\myCARGO})\,,
\end{equation}
where $\Theta(1-\ell z_{\myCARGO})$ denotes the Heaviside step function.
Note that, since the particular expression for $\rho_\circledast(\ell)$ in Eq.~\eqref{eq_BEfracL} diverges for $\ell z_{\myCARGO}\rightarrow1$ and $p(\ell)$ is chosen such that $\ell\leq\ell^*$,
the expression in Eq.~\eqref{eq_rhotBE} is only valid for $\ell z_{\myCARGO}<1$.
Exploiting the mathematical analogy to quantum statistics \cite{huang2009introduction},
we show next how to identify a Bose-Einstein condensation in our model, where $p(\ell)$ takes the role of the density of states.

\subsubsection{Bose-Einstein order parameter}

Regarding the denominator in Eq.~\eqref{eq_BEfracL}, we recognize that
 the fugacity $z_{\myCARGO}$ has an upper bound given by the condition $\ell^*z_{\myCARGO}<1$, which defines the critical fugacity $z_\text{cr}:=1/\ell^*$ (or the critical chemical potential $\mu_\text{cr}:=-k_\text{B}T\ln(\ell^*/\Lambda)$).
Despite this divergence, setting $\ell^*z_{\myCARGO}=1$ is allowed if the limit $z_{\myCARGO}\rightarrow z_\text{cr}$ of Eq.~\eqref{eq_rhotBE} exists, i.e., if the critical density
\begin{equation}\label{eq_rhocrX}
\rho_\text{cr}:=\lim_{z_\myCARGO\rightarrow z_\text{cr}}\rhomyCARGO
\end{equation}
has a finite value.
In this case, a further increase of
the total number density $\rhomyCARGO$ of loaded cargo is possible by loading
the largest carrier, i.e., by increasing the number density $\rho_\circledast(\ell^*)$ at constant $z_\myCARGO=z_\text{cr}$.
While, for $z_\myCARGO<z_\text{cr}$, the expression
\begin{equation}\label{eq_rhoXmax}
\rho_\circledast(\ell^*)=\bar{\rho}_{\myCARRIER}^{\text{t}}\,p(\ell^*)\,\frac{\ell^*z_{\myCARGO}}{1-\ell^* z_{\myCARGO}}\,
\end{equation}
holds, such that the contribution to $\rho_\text{cr}$ is infinitesimal, the value of $\rho_\circledast(\ell^*)$ may become finite as the denominator approaches zero for $z_{\myCARGO}\rightarrow z_\text{cr}$.

In a system with a finite $\rho_\text{cr}$, defined in Eq.~\eqref{eq_rhocrX}, the total number density $\rhomyCARGO$ must account for the explicit contribution of $\rho_\circledast(\ell^*)$.
The resulting generalization of Eq.~\eqref{eq_rhotBE} yields
\begin{equation} \label{eq_rhoNgen}
\rhomyCARGO=\rhoB + \rho_\circledast(\ell^*) +
\frac{\bar{\rho}_{\myCARRIER}^{\text{t}}}{\ell^*}\int_0^{\ell^*}\mathrm{d}\ell\,
\frac{\ell z_{\myCARGO}}{1-\ell z_{\myCARGO}}\,p(\ell)\,\Theta(1-\ell z_{\myCARGO})\,.
\end{equation}
Then, we find that the fraction of cargo particles occupying the largest carrier is given by
\begin{equation}
\label{eq_rhoXBE}
\mathcal{B}:=\frac{\rho_\circledast(\ell^*) }{\rhomyCARGO} = \fu{0}{1- \frac{\rho_\text{cr}}{\rhomyCARGO}} \fuC{if \text{$z_\myCARGO<z_\text{cr}$}\,,}{if \text{$z_\myCARGO=z_\text{cr}$}\,.}
\end{equation}
This ratio constitutes the order parameter of a density-driven Bose-Einstein condensation \cite{huang2009introduction} into a state with a macroscopic occupation of a single carrier species with engulfment strength $\ell^*$ for $\rhomyCARGO>\rho_\text{cr}$.

\subsubsection{Illustration and discussion}

To illustrate the conditions for a Bose-Einstein condensation in  a polydisperse carrier-cargo mixture, we discuss, as a particular example, a system in which all cargo particles (not only those engulfed by a carrier) are non-interacting.
This allows us to express the critical density from Eq.~\eqref{eq_rhocrX} in the closed form
\begin{equation} \label{eq_rhocr}
\rho_\text{cr}=\frac{1}{\ell^*}+\lim_{z_{\myCARGO}\rightarrow z_\text{cr}}\frac{\bar{\rho}_{\myCARRIER}^{\text{t}}}{\ell^*}\int_0^{\ell^*}\mathrm{d}\ell\,
\frac{\ell z_{\myCARGO}}{1-\ell z_{\myCARGO}}\,p(\ell)\,\Theta(1-\ell z_{\myCARGO})\,,
\end{equation}
where we have replaced the number density $\rhoB\rightarrow z_{\myCARGO}$ by the corresponding fugacity, which equals $1/\ell^*$ in the limit taken.

To evaluate the integral in the second term of Eq.~\eqref{eq_rhocr}
we consider a family of normalized distribution functions
\begin{equation} \label{eq_pea}
p(\ell)\propto\left(-\left(\ell-\frac{\ell^*}{2}\right)^2+\frac{\left.\ell^*\right.^2}{4}\right)^\alpha=\ell^\alpha(\ell^*-\ell)^\alpha
\,,
\end{equation}
given by an inverse parabola exponentiated by $\alpha$.
The form of $p(\ell)$ is shown in Fig.~\ref{fig_BEC}a for $\alpha=0,1,2$.
For $\alpha>0$, we have $p(\ell^*)=0$ and the scaling for $\ell\rightarrow\ell^*$ is given by
\begin{equation} \label{eq_peaX}
p(\ell)\propto(\ell^*-\ell)^\alpha+\mathcal{O}\left((\ell^*-\ell)^{\alpha+1}\right)\,.
\end{equation}
 Then $\rho_\text{cr}$ can be explicitly calculated as a function of $\alpha$
and we find that it takes a finite value (blue and green dots in Fig.~\ref{fig_BEC}b and c) if $\alpha>0$, while it diverges logarithmically for $\alpha=0$.

As shown in Fig.~\ref{fig_BEC}b, there are two possibilities for the behavior of the total number density $\rhomyCARGO$ of cargo particles,
depending on the exponent $\alpha$ of $p(\ell)$ as $\ell\rightarrow\ell^*$ and thus on the critical density $\rho_\text{cr}$.
First, if $\rho_\text{cr}$ diverges, all carriers take up an infinitesimal fraction of cargo particles for $z_{\myCARGO}<z_\text{cr}$ and no phase transition occurs.
Second, if $\rho_\text{cr}$ remains finite, then $\rhomyCARGO$ can be increased indefinitely at constant critical fugacity $z_{\myCARGO}= z_\text{cr}$ by loading further cargo particles on the carriers with largest engulfment strength $\ell^*$.
As a consequence, a Bose-Einstein condensation occurs at $\rhomyCARGO=\rho_\text{cr}$ towards a state with a macroscopic number of cargo occupying the carriers with $\ell=\ell^*$.

For $z_{\myCARGO}<z_\text{cr}$, the behavior of $\rhomyCARGO$ as a function of $z_{\myCARGO}$ is given by either Eq.~\eqref{eq_rhotBE} or Eq.~\eqref{eq_rhoNgen}, since $\rho_\circledast(\ell^*)=0$.
Therefore, the order parameter $\mathcal{B}$ in Eq.~\eqref{eq_rhoXBE} remains zero.
However, when $\rhomyCARGO=\rho_\text{cr}$ is exceeded for $z_{\myCARGO}=z_\text{cr}$, we have $\rho_\circledast(\ell^*)>0$,
Hence, the value of $\mathcal{B}$ increases continuously as a function of $\rhomyCARGO$ in the new phase.
This behavior is illustrated in Fig.~\ref{fig_BEC}c.

We conclude that there is no Bose-Einstein condensation for a constant distribution of engulfment strengths in a certain interval, while the formation of a Bose-Einstein condensate is facilitated for mixtures of carriers with an increasingly weaker polydispersity (such that $p(\ell)$ has a smaller variance).
This observation is analogous to Bose-Einstein condensation in quantum statistics, where the density of states of a three dimensional ideal Bose gas has the exponent $1/2$ and gives rise to a phase transition, in contrast to the exponents $0$ and $-1/2$ in two and one dimensions, respectively.
Finally, we stress that the crucial technical difference between a classical system and quantum statistics is that, in the present case, the momenta of the particles are thermalized and represented by the thermal wave length $\Lambda$.
Thus, following the discussion in Sec.~\ref{sec_attraction}, the respective energy levels $E:=-k_\text{B}T\ln(\ell/\Lambda)$ are provided by a configurational quantity modeling the non-equilibrium drive of the carriers to engulf their cargo.
Moreover, the spatial dimension (taken here as $d=1$) would only enter here as a trivial factor.

\section{Conclusions \label{sec_conclusions}}

In this work, we have introduced a combination law~\eqref{eq_geneff} that illustrates that the internal degrees of freedom associated with the formation of soft-matter complexes can be exactly mapped onto effective chemical potentials, which allow to describe such complexes as an independent particle species.
We demonstrated the validity of our approach for carrier-cargo mixtures through
exactly matching simulation results for a one-dimensional hard-rod model, where the formation of CCCs is purely driven by entropic free-volume effects.
Then we argued that non-Boltzmann occupation laws could provide a heuristic description of non-equilibrium particle uptake, as exemplified by recognizing a Bose-Einstein condensation for a polydisperse carrier model.

The versatility of our approach
can be exploited when accounting for emergent properties of the assembled complexes that are distinct from their building blocks. While the present application to carrier-cargo mixtures involves CCCs whose external length
is strictly set by the carrier, we can also account for a shape change upon cargo uptake.
Likewise, by appropriately adapting the interactions, it is possible to describe various types of aggregates~\cite{sacanna2013shaping} such as
dimers formed by lock and key colloids \cite{sacanna2010lock,sacanna2011lock,wang2014lock,kamp2016electric},
colloidal molecules \cite{wang2012colloids,li2011colloidal,duguet2011design,kogler2015generic,wan2021polymerizationcolloidalmolecules,lowen2018activecolloidalmolecules}
or entropic bonds \cite{harper2019entropic,vo2022entropicbondtheory} with possible applications for characterizing orientational order in DFT~\cite{martinez2021failure,martinez2022effect}.
Exploiting the full potential of modern fundamental-measure DFT \cite{hansen2006WBII,roth2010fundamental,roth2012,wittmann2016,wittmann2017phase} the statistical treatment of such anisotropic interactions or inhomogeneities induced by external stimuli is within reach.

Another important application of our model would be a dynamical description of cargo uptake,
 which can be achieved in two ways.
The first possibility is via direct computer simulation of the dynamics. This requires, however, soft models for the carrier membrane as discussed in Sec.~\ref{sec_soft}.
The second possibility amounts to a direct application of our effective picture
through an extension \cite{lutsko2016mechanism,moncho2020scaling} of dynamical DFT \cite{marconi1999DDFT,archer2004DDFT,tevrugt2020revDDFT}
in which appropriate reaction rates should be determined from the equilibrium statistics.
Such dynamical approaches could also be used to explicitly study more complex carrier-cargo mixtures involving self-propelled building blocks \cite{lowen2018activecolloidalmolecules,kunti2021rational,vutukuri2020active,peterson2021vesicle,popescu2011pulling,baraban2012transport,vuijk2021,muzzeddu2023taxis,willemsalvarez2022prep}
or emerging active motion of the assembled complex \cite{palacci2013photoactivated,martinez2015colloidal,niu2018dynamics,schmidt2019light,grauer2021droploids,alvarez2021reconfigurable}.
Moreover, it would be interesting to simulate an explicit microscopic model with non-equilibrium particle uptake
for testing the heuristic idea of non-Boltzmann occupation laws and for exploring the possibility to find an appropriate coarse-graining scheme.

 A critical open question which could stimulate future efforts concerns the experimental realization of cargo uptake and release in equilibrium soft-matter systems, e.g., involving colloidal carriers or lipid vesicles.
Further experimental challenges will be to discover or train functional carriers \cite{odziomek2012conception} whose non-equilibrium driving force for cargo uptake is not limited to Boltzmann occupation
or develop intelligent micromachines \cite{huang2022micromachines} which learn to collect their cargo as programmed.
In regard of this perspective, it would be interesting to test these ideas with an explicit microscopic model and also explore the possibility to find an appropriate coarse-graining scheme.

\acknowledgements

The authors would like to thank Michael te Vrugt, Roland Roth, Daniel Borgis, Laura Alvarez and Philipp Maass for stimulating discussions and valuable suggestions.
RW and HL acknowledge support by the Deutsche Forschungsgemeinschaft (DFG) through the SPP 2265, under grant numbers WI 5527/1-1 (RW) and LO 418/25-1 (HL).

\appendix

\begin{figure}[t]
\begin{center}
\includegraphics[width=0.95\linewidth]{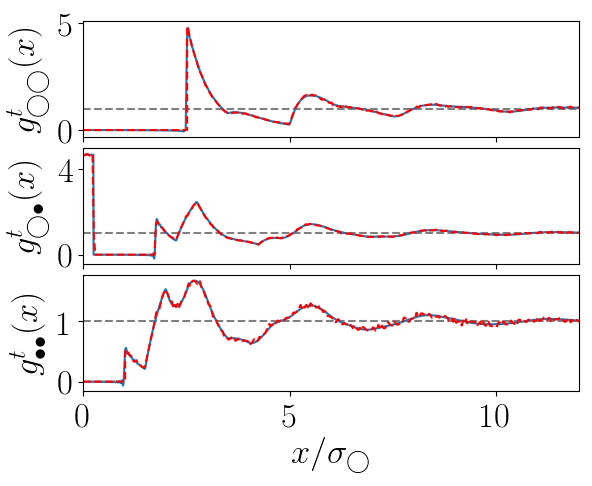}
\caption{\label{fig_g_simulation}
 Excellent agreement of simulation (dashed red) and theory (solid blue) for the physical pair distributions ${g}^\text{t}_{ij}(x)$ of a hard-body carrier-cargo mixture with one possible CCC in $d=1$ dimension, as discussed in Sec.~\ref{sec_resultsk1STRUCTURE}.
The parameters $\sigma_\myCARGO=0.4\sigma_\myCARRIER$ and $\lambda_\myCARRIER=0.6\sigma_\myCARRIER$ and $\rhomyCARGO\sigma_\myCARRIER=0.625$ and  $\rhomyCARRIER\sigma_\myCARRIER=0.72$ are the same as in Fig.~\ref{fig_CCCfractions123dp2},
such that the theory results correspond to the bottom panel.}
\end{center}
\end{figure}

\section{\label{app_MC} Monte-Carlo simulation}

To corroborate our analytic treatment, we perform canonical Monte-Carlo simulation of bulk systems of $M$ carriers and $N$ cargo particles, realized via periodic boundary conditions, in $d=1,2,3$ dimensions.
 The total particle number $M+N=2000$ is always fixed, while we vary the ratio $N/M=\rhomyCARGO/\rhomyCARRIER$ of cargo to carrier particles between independent simulation runs.
The input densities $\rhomyCARGO$ and $\rhomyCARRIER$ are determined by the final size of the simulation box.
The overlap criterion between a pair of particles is given by Eqs.~\eqref{eq_Uijrod1} and \eqref{eq_Uijrod2}.
We perform $10^6$ Monte Carlo cycles, each consisting of $M+N$ trial moves.
In each trial move, one of the particles is displaced by a distance vector $\Delta\mathbf{r}$, whose components $\Delta r_i$ with $i=1,\ldots,d$ are independent random numbers drawn from a uniform distribution in the interval $\left[-\Delta_{\mathrm{max}},\Delta_{\mathrm{max}}\right]$.
Any trial move is accepted with probability $P=\min(1,\exp(-{\Delta U}/{k_\text{B} T}))$ \cite{metropolis}, i.e, overlap free trial moves are accepted, while overlapping trial moves are rejected.
 Over the course of the simulation, the value of
$\Delta_{\mathrm{max}}$ is adjusted such that acceptance ratio remains close to $0.3$.

We obtain the particle configurations at high densities by following a compression protocol. We randomly initialize the system at low number densities $\rhomyCARGO + \rhomyCARRIER = 5 \times 10^{-3} /  \sigma_\myCARRIER$.
Over the course of the simulation, we gradually rescale all coordinate axes $x_i, i \in \{1,2,3\}$, such that $x_i \propto \tau^{-1/(3d)}$, where $d$ is the number of spatial dimensions, effectively increasing all total densities within the system as $\rhomyCARGO \propto \rhomyCARRIER \propto \tau^{1/3}$. Here,
$\tau \in [0,1] $ denotes the simulation progress, i.e., the number of completed Monte-Carlo cycles.
This type of decelerating compression aids the equilibration speed, since the system is quickly compressed in the density regime where the particles are expected to rarely come in contact, while being allowed to undergo a larger fraction of Monte Carlo cycles at higher densities.

Exemplary simulation data are shown as dots in Fig.~\ref{fig_CCCfractions123dp1}.
In addition, Fig.~\ref{fig_g_simulation} shows the pair distribution functions in one dimension for $\rhomyCARGO\sigma_\myCARRIER=0.625$ and $\rhomyCARRIER\sigma_\myCARRIER=0.72$. Additionally, corresponding results are shown, obtained from the average over $200$ simulation runs (dashed red).
As visible, the theoretical graphs (blue, cf.\ Fig.~\ref{fig_CCCfractions123dp2}), calculated as described in Sec.~\ref{sec_pairdistributions}, agree with the simulation results to an excellent degree.

\section{Canonical partition function of a hard-rod mixture in one dimension \label{app_Z}}

To aid the discussion in Sec.~\ref{sec_c3}, we state here the exact canonical partition function
\begin{align}
\!\!\!\!\!\!&\mathcal{Z}^{(L)}_{N_1,N_2,\ldots, N_K}(\sigma_1,\sigma_2,\ldots,\sigma_K)\cr&=\prod_{i=1}^K \frac{1}{N_i!\Lambda^{N_i}}\left(L-\sum_{j=1}^K N_j\sigma_j\right)^{N_i}\!\!\! \Theta\!\left(L-\sum_{j=1}^K N_j\sigma_j\right)\ \ \ \ \ \ \
\label{eq_Zkappa}
\end{align}
 a $K$-component mixture of hard rods (one-dimensional hard spheres or Tonks gas).
 Specifically, $N_i$ denote the particle numbers of each species $i\in\{1,2,\ldots,K\}$ confined in a one-dimensional interval of length $L$, while we ignore the trivial dependence on the temperature $T$ through the thermal wavelength $\Lambda$.
For later convenience, we further write $\mathcal{Z}^{(L)}_{N_1,N_2,\ldots, N_K}$ as a function of the particle lengths $\sigma_i$ of each species, while the explicit dependences on the particle numbers and the system length are written as a subscript and a superscript, respectively.
The Heaviside function $\Theta(x)$ merely ensures that the system is below close packing for the given particle numbers and is silently omitted elsewhere to ease the notation.

\section{Canonical partition function of the carrier-cargo mixture \label{app_Z1dfact}}

To better understand the notion of our system consisting of $M$ carriers and $N$ cargo as an effective three-component mixture, we rewrite Eq.~\eqref{eq_ZMN} in the less intuitive but more instructive form
\begin{align}
\label{eq_ZMNi}
Z_{M,N} =\sum_{C=0}^{\min(M,N)} \mathcal{Z}^{(L)}_{M-C,N-C,C}(\sigma_{\myCARRIER},\sigma_{\myCARGO},\sigma_{\myCARRIER})\left(\mathcal{Z}^{(\lambda_{\myCARRIER})}_1(\sigma_{\myCARGO})\right)^C\,,
\end{align}
which contains no factorial expressions.
In the second equality, we made use of the fact that the free length
\begin{align}
\ell\equiv \Lambda \mathcal{Z}^{(\lambda_{\myCARRIER})}_1(\sigma_{\myCARGO})\,.
\label{eq_epsilonentropic1}
\end{align}
accessible to a cargo particle within a CCC, as defined in Eq.~\eqref{eq_epsilondef},
can be explicitly interpreted as the canonical partition function
of single particle, i.e., Eq.~\eqref{eq_Zkappa} with $K=1$ and $N_1=1$.
Therefore, we see that the total partition function $Z_{M,N}$ of our carrier-cargo mixture can be perceived as sum over products of different partition functions: $\mathcal{Z}^{(L)}_{M-C,N-C,C}$ represents the external interactions and $\mathcal{Z}^{(\lambda_{\myCARRIER})}_1$ denotes the internal occupation statistics of a single CCC.
By external interaction we generally mean the physical interactions between the particles excluding the possibility to engulf one another, which here is the hard-core repulsion of diameters $\sigma_{\myCARRIER}$ and $\sigma_{\myCARGO}$.
In other words, we have demonstrated that
an occupied carrier represents a third species of size $\sigma_{\myCCC}:=\sigma_{\myCARRIER}$ which additionally possesses internal degrees of freedom.
We can therefore speak of an effective mixture of $A:=M-C$ empty carriers, $B:=N-C$ free cargo particles and $C$ CCCs (occupied carriers).

\section{Relation between fugacities and homogeneous number densities \label{app_ideal}}

In Eq.~\eqref{eq_Xi} we have derived the grand partition function for an effective ternary system representing the carrier-cargo mixture.
To prepare for the effective treatment of more general mixtures,
it is instructive to discuss an even simpler mapping.
In fact, since empty carriers have the same length $\sigma_{\myCARRIER}\equiv\sigma_{\myCCC}$ as the corresponding CCCs,
a two-species picture is sufficient if one is only interested in information contained in $\rhomyCARRIER$.
To see this, we recast Eq.~\eqref{eq_Xi} as
\begin{align}\label{eq_Xitest}
\Xi&=\sum_{M=0}^\infty\sum_{B=0}^\infty \frac{(z_{\myCARRIER}+{z_{\myCCC}})^Mz_{\myCARGO}^B}{M!B!}\left(L-M\sigma_{\myCARRIER}-B\sigma_{\myCARGO}\right)^{(M+B)}\,
\end{align}
by undoing the substitution $M \rightarrow A+C$ and rearranging the sums to obtain a
binomial series. In such a two-species mixture, the carrier species (with no distinction of empty or occupied)
can be interpreted to be coupled to a single effective particle reservoir
with an enhanced fugacity $z_{\myCARRIER}^\text{t}:=z_{\myCARRIER}+{z_{\myCCC}}$ also  accounting for the internal degrees of freedom.
Thus, Eq.~\eqref{eq_Xitest}
illustrates the additivity $\rhomyCARRIER=\rhoA+\rho_\myCCC$ of the effective densities in Eq.~\eqref{eq_densities} directly by the additivity of the corresponding fugacities.

More generally, as long as the interactions $\sigma_{\myCARRIER}=\sigma_{\myCCC}$ between two (or more) species are the same,
   one finds the following scaling relation between densities and fugacities:
  \begin{align}\label{eq_scalingdensfug}
\frac{\rhomyCARRIER}{z_{\myCARRIER}^\text{t}}=\frac{\rhoA}{z_{\myCARRIER}}=\frac{\rho_\myCCC}{z_{\myCCC}}\,.
\end{align}
We thus see that the fraction $\rho_\myCCC/\rhomyCARRIER$ of CCCs (the number of carriers occupied by a cargo divided by the total number of carriers) can be determined solely from the corresponding fugacities for any interaction between the particles as $\rho_\myCCC/\rhomyCARRIER=z_{\myCCC}/z_{\myCARRIER}^\text{t}$.
To determine the CCC fraction as an explicit function of total carrier and cargo density, as, e.g., in Fig.~\ref{fig_CCCfractions123dp1}, further calculations are necessary.

The above relations between (effective) fugacities and number densities become most apparent in an effective system without interactions, i.e., when setting $\sigma_{\myCARRIER}=\sigma_{\myCARGO}=\sigma_{\myCCC}=0$.
In this ideal case, indicated by the superscript (id), the homogeneous number densities $\rhoA^\text{(id)}=z_{\myCARRIER}$  and  $\rho_\myCCC^\text{(id)}=z_{\myCCC}$
are explicitly given by the corresponding fugacities.
Moreover, without (external) interactions, it is easy to see from either Eq.~\eqref{eq_Xi} or Eq.~\eqref{eq_Xitest} that the grand partition function can be
explicitly written in the closed form
\begin{align}\label{eq_Xi3specidsum}
\Xi^\text{(id)}&=e^{L(z_{\myCARRIER}+z_{\myCARGO}+{z_{\myCCC}})}\,.
\end{align}
This result underlines that the combinatorics underlying the generic definition, Eq.~\eqref{eq_3spec}, of the effective fugacity ${z_{\myCCC}}$, is independent of the explicit interactions.

For practical reasons, considering two species with fugacities $z_{\myCARRIER}^\text{t}$ and $z_{\myCARGO}$ in Eq.~\eqref{eq_Xitest} is not very helpful, since the
information on the distribution $\rho_\myCCC$ of CCCs and thus also the total density $\rhomyCARGO$ of cargo particles is not available, as it requires knowing both $\rhoA$ and $\rhomyCARRIER$.
As such, it is in general more appropriate to work with Eq.~\eqref{eq_Xi} and to distinguish between physically indistinguishable particles (empty carriers and CCCs) in our effective system.
Then, $\rhomyCARGO$ can be determined from the auxiliary relation $\rhomyCARGO=\rhoB+\rho_\myCCC$ in Eq.~\eqref{eq_densities}.
However, it is not possible to derive $\rhomyCARGO$ in the spirit of Eq.~\eqref{eq_Xitest}, i.e., directly from the statistics underlying $\Xi$. This is because loaded cargo does not contribute to external interactions.
In other words, the total packing fraction $\eta_\myCARRIER^\text{t}:=\rhomyCARRIER\sigma_\myCARRIER=\rhoA\sigma_\myCARRIER+\rho_\myCCC\sigma_\myCARRIER$ of carriers is not affected by the number of CCCs and follows the addition law, Eq.~\eqref{eq_densities}, of number densities, while only the free cargo particles contribute to the total packing fraction $\eta_\myCARGO^\text{t}:=\rhoB\sigma_\myCARGO\neq\rhomyCARGO\sigma_\myCARGO$.

\section{Classical Density Functional Theory (DFT) \label{app_DFT}}

Here we provide a compact introduction to classical DFT \cite{Evans1979,hansenmcdonald}, a powerful toolbox to determine the configuration and structure of interacting fluid mixtures. First, in Sec.~\ref{sec_DFT}, we introduce the DFT framework for general inhomogeneous fluid mixtures and explain how to determine the homogeneous number densities in our effective description of carrier-cargo mixtures.
Second, in Sec.~\ref{sec_geff}, we explain how to determine the effective pair distribution functions, from which we later recover in Appendix~\ref{app_gtrue} the exact expressions for the physical carrier-cargo mixture
in one spatial dimension.

\subsection{Composition of a mixture from DFT\label{sec_DFT}}

Consider in general a fluid mixture of $K$ different components in the external one-body potentials $V^{(i)}_\text{ext}(\mathbf{r})$ acting on the particles of species $i\in\{1,2,\ldots,K\}$.
Then, the corresponding (inhomogeneous) number densities $\rho_i(\mathbf{r})$ in equilibrium can be obtained from solving the Euler-Lagrange equations
\begin{align}\label{eq_ELS}
\frac{\delta\Omega[\{\rho_i\}]}{\delta\rho_i(\mathbf{r})}=0\,,
\end{align}
where $\Omega[\{\rho_i\}]$ is a density functional which becomes minimal when the equilibrium solutions are inserted.
This minimal value of the functional corresponds to the grand potential $\beta\Omega=-\ln\Xi$ of the system.
Hence, we can recover the grand-canonical partition functions $\Xi$ calculated in the main text (or appropriate generalizations to arbitrary external fields).
As a prerequisite, we need to know the proper functional.

The general form of the density functional reads
\begin{align}\label{eq_OM}
\Omega[\{\rho_i\}]=\Omega^\text{(id)}[\{\rho_i\}]+\mathcal{F}_\text{ex}[\{\rho_i\}]\,,
\end{align}
where the excess (over ideal gas) free energy $\mathcal{F}_\text{ex}$ describes the interactions between the particles and
the exactly known functional $\Omega^\text{(id)}$ for
 an ideal gas of point-like particles in an externally applied potential $V^{(i)}_\text{ext}(\mathbf{r})$ reads
\begin{align}\label{eq_OMid}
\!\!\!\beta\Omega^\text{(id)}=\sum_{i=1}^K\int\mathrm{d}\mathbf{r}\,\rho_i(\mathbf{r})\left(\ln\!\left(\frac{\rho_i(\mathbf{r})}{z_i}\right)-1+\beta V^{(i)}_\text{ext}(\mathbf{r})\!\right)\!\!\!
\end{align}
in $d$ spatial dimensions,
recalling the definitions $\beta=(k_\text{B}T)^{-1}$ of the inverse temperature and
 $z_i:=\exp(\beta\mu_i)/\Lambda^d$ of the fugacities.

For the hard-body interactions considered in this work, we employ fundamental measure theory (FMT) \cite{hansen2006WBII,roth2010fundamental,roth2012} for the excess free energy $\mathcal{F}_\text{ex}$ which follows the same recipe in all spatial dimensions.
Specifically, the FMT in one spatial dimension \cite{percus1976equilibrium,vanderlickpercus1989mixture}, most commonly known as the Percus functional, is exact.
Here,
\begin{equation}
\beta\mathcal{F}_{\text{ex}}=-\int\mathrm{d} x\,n_0(x)\ln (1-n_1(x))
\end{equation}
follows as a function of the weighted densities
\begin{equation}
  n_{0/1}(x) = \sum_{i=1}^{K} \int \mathrm{d} x_1 \,  \rho_i(x_1)\, \omega^{(0/1)}_{i}(x-x_1)
\end{equation}
which consist of convolution integrals
of the densities and the weight functions
\begin{align}
\omega^{(1)}_i(x)&=\Theta(\sigma_i/2-|x|)\,,\cr
\omega^{(0)}_i(x)&=\frac{1}{2}(\delta(\sigma_i/2-x)+\delta(\sigma_i/2+x))\,,
\end{align}
where $\sigma_i$ is the length of a hard rod of species $i$.
These weight functions represent the geometry of particle $i$ being local measures of the one-dimensional volume (rod length) and the surface area (characteristic function), respectively.
In higher spatial dimensions, there exists a larger set of required weight functions.

For our calculations we consider homogeneous bulk systems with $V^{(i)}_\text{ext}(\mathbf{r})=0$,
such that the number densities $\rho_i$ do not depend on the position $\mathbf{r}$.
Then, the one-dimensional weighted densities simply read
\begin{align}
n_0=\sum_{i=1}^{K}\rho_i\,,\ \ \ \ \ \
n_1=\sum_{i=1}^{K}\rho_i\sigma_i \,,
\label{eq_wds}
\end{align}
such that the functional $\Omega[\{\rho_i\}]$ turns into an explicit function of the number densities and the functional derivative in Eq.~\eqref{eq_ELS} turns into a partial derivative.
Thus, we only need to solve a set of algebraic equations to find the desired relation between the fugacities and densities.
In one  spatial dimension these are
\begin{align}
\ln(\rho_i/z_i)=-\sum_{\nu=1}^2\frac{\partial \mathcal{F}_\text{ex}}{\partial n_\nu}\frac{\partial n_\nu}{\partial\rho_i}=\ln(1-n_1)-\frac{n_0}{1-n_1}\sigma_i
\label{eq_ELGhom}
\end{align}
for $i=1,\ldots,K$.
In higher spatial dimensions, the structure is exactly the same but then the sum over $\nu$ must include the additional weighted densities.

To briefly connect to our previous statistical results, let us note that upon applying the variational scheme from Eq.~\eqref{eq_ELS} to an ideal gas in an external field, with the functional from Eq.~\eqref{eq_OMid}, it is easy to show that the partition function reads
\begin{align}\label{eq_XiGENext}
\Xi^\text{(id)}&=
\exp\left(\sum_{i=1}^K \left(\int\mathrm{d}\mathbf{r}\,e^{-\beta V^{(i)}_\text{ext}(\mathbf{r})}\right)z_i\right)\,.
\end{align}
In the absence of an external potential the integral must be replaced by the system volume  $L^d=\int\mathrm{d}\mathbf{r}$.
We have thus recovered Eq.~\eqref{eq_XiGEN} from our DFT formalism upon setting $K=k+\kappa$ and using the corresponding (effective) fugacities.
Likewise, with interactions, we can also recover the exact result of Eq.~\eqref{eq_Xi}
for the particular one-dimensional case,
while the DFT calculation directly extends (in a good approximation) to higher spatial dimensions.

\subsection{Structure of a mixture from DFT \label{sec_geff}}

Structural information on the system can be extracted from DFT by calculating functional derivatives of the excess free energy.
In particular, the direct correlation functions $c_{ij}(r)$ of a fluid mixture with $i,j\in\{1,2,\ldots,K\}$ are defined as
\begin{equation}
c_{ij}(r=|\mathbf{r}_1-\mathbf{r}_2|)=-\frac{\delta^2\beta\mathcal{F}_\text{ex}}{\delta \rho_i(\mathbf{r}_1)\delta \rho_j(\mathbf{r}_2)}\,.
\label{eq_DCFdft}
\end{equation}
For a homogeneous system of hard rods, we find the exact direct correlation functions
\begin{align}\label{eq_DCF}
\!\!\!c_{ij}(x)=-\frac{n_0}{(1-n_1)^2}\,W_{ij}^{(11)}(x)
-\frac{1}{1-n_1}\,W_{ij}^{(10)}(x)\!\!\!
  \end{align}
 from the Percus functional with the weighted densities given by Eq.~\eqref{eq_wds} and the functions
\begin{align}
W_{ij}^{(11)}(x)&= \min(\sigma_i,\sigma_j)\,\Theta(\Delta a-|x|)\cr&\ \ \ +(a_{ij}-|x|)\,\Theta(a_{ij}-|x|)\,\Theta(|x|-\Delta a)\,,\cr
W_{ij}^{(10)}(x)&= \Theta(a_{ij}-|x|)\,,\ \ \
\end{align}
where $a_{ij}:=(\sigma_i+\sigma_j)/2$, $\Delta a:=|\sigma_i-\sigma_j|/2$ and $\min(\sigma_i,\sigma_j)$ returns the smaller one of the two length.
In general, $W_{ij}^{(\nu\mu)}$ can be calculated through convolution products of $\omega_i^{(\nu)}$ and $\omega_j^{(\mu)}$.
Note that, if $i=j$, we have $\Delta a=0$ and only the second term in $W_{ii}^{(11)}(x)$ is relevant.

Next, the pair distributions ${g}_{ij}(x)=h_{ij}(x)+1$ can be calculated
from the total correlation functions $h_{ij}(x)$,
which are related to the direct correlation functions $c_{ij}(x)$ via the multicomponent Ornstein-Zernike equation
\begin{align}
    h_{ij}(x_1-x_2)&=c_{ij}(x_1-x_2)\\\nonumber
    & \ \  +\sum_{l=1}^{K}\rho_l\int\mathrm{d} x_3\, c_{il}(x_1-x_3)\,h_{lj}(x_3-x_2)\,.
    \label{eq_OZ}
\end{align}
The matrix solutions in Fourier space read
\begin{align}
\hat{H}(q)=(\boldsymbol{1}-\hat{C}(q))^{-1}-\boldsymbol{1}\,,
\end{align}
where the components of the auxiliary matrices $\hat{H}_{ij}(q)=\sqrt{\rho_i\rho_j}\hat{h}_{ij}(q)$ and
$\hat{C}_{ij}(q)=\sqrt{\rho_i\rho_j}\hat{c}_{ij}(q)$ can be obtained by weighting the Fourier transforms $\hat{h}_{ij}(q)$ and $\hat{c}_{ij}(q)$ of $h_{ij}(x)$ and $c_{ij}(x)$, respectively, with the corresponding homogeneous densities.
For hard rods in one dimension it is possible to determine an exact analytical solution of Eq.~\eqref{eq_OZ} \cite{santos2007exact}.
However, for simplicity, we perform the final inverse Fourier transform of $\hat{h}_{ij}(q)$ numerically.

\section{Pair distributions for a one-dimensional carrier-cargo mixture \label{app_gtrue}}

In this Appendix, we provide further background on the relations, stated in Sec.~\ref{sec_pairdistributions}, between the effective and physical pair distributions of the carrier-cargo mixture.
For our minimal model of a carrier holding at most one cargo, the basic relation
between the effective number densities, $\rhoA$, $\rhoB$ and $\rho_\myCCC$,
and the total densities, $\rhomyCARRIER$ and $\rhomyCARGO$,
of the physical system is given by Eq.~\eqref{eq_densities}.
For the pair distributions these relations are generally more involved since a CCC represents both carrier and cargo.

To determine the effective pair distributions ${g}_{ij}(x)$ with $i,j\in\{\myCARRIER,\myCARGO,\myCCC\}$ for the mapped hard-rod system in one dimension
and the pair distributions ${g}^\text{t}_{ij}(x)$ with $i,j\in\{\myCARRIER,\myCARGO\}$ for our physical carrier-cargo mixture,
let us first recall from Appendix~\ref{app_DFT} that the former follow directly from the exact direct correlation function, Eq.~\eqref{eq_DCF}, in the effective system.
Moreover, following the structural equivalence of empty carriers and CCCs (both species have the same hard-core diameter $\sigma_\myCARRIER$) we have
\begin{align}
g_{\myCARRIER\myCARRIER}\equiv g_{\myCARRIER\myCCC}\equiv g_{\myCCC\myCCC}\,.
\end{align}
As a result, it is sufficient to determine the pair distribution
for an effective binary mixture of hard rods with $\rhomyCARRIER=\rhoA+\rho_\myCCC$ and $\rhoB$, as discussed in Appendix~\ref{app_ideal}.
For this reason, we apparently have $g^\text{t}_{\myCARRIER\myCARRIER}=g_{\myCARRIER\myCARRIER}$ in Eq.~\eqref{eq_gactA}.
The remaining pair distributions of the physical system require additional care.

To understand the formulas in Sec.~\ref{sec_pairdistributions}, we first point out that the relation, Eq.~\eqref{eq_densities}, between the one-body densities
analogously applies to the two-body densities $\rho_i\rho_j{g}_{ij}$ and $\rho^\text{t}_i\rho^\text{t}_j{g}^\text{t}_{ij}$.
Solving for ${g}^\text{t}_{ij}$ yields the basic structure of Eqs.~\eqref{eq_gactA} to~\eqref{eq_gactD}
Specifically, $g^\text{t}_{\myCARRIER\myCARRIER}=g_{\myCARRIER\myCARRIER}$, directly follows as
\begin{align}
g^\text{t}_{\myCARRIER\myCARRIER}&=\frac{\rhoA^2g_{\myCARRIER\myCARRIER}+\rhoA\rho_\myCCC(g_{\myCARRIER\myCCC}+g_{\myCCC\myCARRIER})+\rho_\myCCC^2g_{\myCCC\myCCC}}{(\rhomyCARRIER)^2}\,.
\end{align}
Second, for the remaining pair distributions, we need to take into account
the fact that the center of a cargo particle, which is contained in a CCC
does not necessarily coincide with the center of the carrier
but is uniformly distributed within the accessible space of length $\ell$.
Therefore, the effective distributions need to be blurred by calculating a
convolution with the indicator function $\Theta(\ell/2-|x|)$ in the corresponding coordinate(s).
This is indicated by the superscript (b) in Eqs.~\eqref{eq_gb} and~\eqref{eq_gbb}.
Third, specifically for the correlations between carrier and cargo particles in Eqs.~\eqref{eq_gactC} and~\eqref{eq_gactD},
we need to manually account for the presence of a cargo particle within a carrier.
This is achieved by redefining the effective two-body density $\rho_{\myCCC\myCCC}^{(2)}$ upon adding this self contribution (which for an ordinary fluid must be subtracted from the two-point correlation function to recover the two-body density).
In practice, this amounts to setting $\rho_{\myCCC\myCCC}^{(2)}(x)\rightarrow \rho_{\myCCC\myCCC}^{(2)}(x)+\rho_\myCCC\delta(x)$,
and then blurring the whole expression as described above.
This yields the generalized contribution
\begin{align}
G^\text{(b)}_{\myCCC\myCCC}(x_1-x_2)&:=\frac{1}{\ell}\int\mathrm{d}x' \Theta(\ell/2-|x_2-x'|)\\\nonumber
&\ \ \ \ \times \left(\rho_\myCCC^2 g_{\myCCC\myCCC}(|x_1-x'|)+\rho_\myCCC\delta(x_1-x')\right),
\end{align}
which can be simplified to yield Eq.~\eqref{eq_Gbb}.

\section{Mixtures with multiple carriers \label{app_multiplecarriers}}

The relations from Sec.~\ref{sec_combiCC} can be easily generalized to describe mixtures involving cargo with fugacity $z_{\myCARGO}$ and now $\kappa-1$ different carrier species with fugacities $z_{\myCARRIER}^{[i]}$ for $i\in\{1,\ldots,\kappa-1\}$.
In particular, introducing for each carrier the typical lengths $\ell^{[i]}$ entering in the $\ell_\nu^{[i]}$ (compare, e.g., Eq.~\eqref{eq_epsilonentropicGEN}), we can define as in Eq.~\eqref{eq_geneffMU} the effective fugacities $z_{\myCCC}^{[i,\nu]}$ of a CCC composed of one carrier of species $i$ and exactly $\nu$ cargo.
Thus, Eq.~\eqref{eq_geneffMUsum} generalizes to
 \begin{equation}
    z_{\myCCC}^{[i]} := \sum_{\nu=1}^\infty z_{\myCCC}^{[i,\nu]}=
    z_{\myCARRIER}^{[i]} \sum_{\nu=1}^\infty \left(\left(\ell_\nu^{[i]}\right)^d z_{\myCARGO}\right)^\nu\,.
    \label{eq_geneffMUsum_MIX}
\end{equation}
By accordingly rewriting Eqs.~\eqref{eq_rhoMN_inf_CCCfrac} and \eqref{eq_rhoMN_inf_LCfrac} as
\begin{equation}\label{eq_rhoMN_inf_MIX0}
\rho_\myCCC^{[i]}:=\sum_{\nu=1}^\infty\rho_{\myCCC}^{[i,\nu]}\,,\ \ \ \rho_\circledast^{[i]}:=\sum_{\nu=1}^\infty\nu\rho_{\myCCC}^{[i,\nu]} \,
\end{equation}
for each carrier, we obtain from Eq.~\eqref{eq_rhoMN_gen} the total number densities
\begin{align}\label{eq_rhoMN_inf_MIXa}
\rhomyCARRIER=
\sum_{i=1}^{\kappa-1}\rho_{\myCARRIER}^{([i],\text{t})}\,, \\
\rho_{\myCARRIER}^{([i],\text{t})}=\rhoA^{[i]}+\rho_\myCCC^{[i]}\,,\label{eq_rhoMN_inf_MIXb} \\
\rhomyCARGO=
\rhoB+\sum_{i=1}^{\kappa-1}\rho_\circledast^{[i]} \,.\label{eq_rhoMN_inf_MIXc}
\end{align}
As an application, the results for the CCC fraction and loaded-carrier fraction of species $i$ can be directly obtained for the respective occupation laws from Eqs.~\eqref{eq_fractionsBO},~\eqref{eq_fractionsFD} or~\eqref{eq_fractionsBE} by simply introducing the superscript $[i]$ to $\ell\rightarrow\ell^{[i]}$ as a species label.

As a next step, let us consider the case $\kappa\rightarrow\infty$ of a polydisperse mixture of carriers,
whose engulfment strengths $\ell^{[i]}\rightarrow\ell$ are continuously distributed according to a normalized distribution $p(\ell)$, such that all sums
\begin{align}\label{eq_sumint}
\sum_{i=1}^{\kappa-1}f^{[i]}\rightarrow\frac{1}{\ell^*}\int_0^{\ell^*}\mathrm{d}\ell \,f(\ell)
\end{align}
turn into integrals and the superscript $[i]$ of a quantity $f^{[i]}$ turns into the argument $\ell$ of a function $f(\ell)$.
By introducing the upper bound $\ell^*$ in Eq.~\eqref{eq_sumint} we imply that the distribution $p(\ell>\ell^*)=0$ vanishes beyond a maximal size $\ell^*$ to prevent a collapse towards (uncontrolled) infinite occupation for $\ell\rightarrow\infty$.
Specifically, let us prescribe the function $p(\ell)$ such that the total number density $\rhomyCARRIER(\ell)=\bar{\rho}_\myCARRIER^\text{t} p(\ell)$ of each carrier with engulfment strengths $\ell$ is a specified input quantity.
Then we can recast Eq.~\eqref{eq_rhoMN_inf_MIXa} in the continuous form
\begin{align}
\bar{\rho}_\myCARRIER^\text{t}=
\frac{1}{\ell^*}\int_0^{\ell^*}\mathrm{d}\ell \,\rhomyCARRIER(\ell)=
\frac{\bar{\rho}_\myCARRIER^\text{t}}{\ell^*}\int_0^{\ell^*}\mathrm{d}\ell \,p(\ell)\,,
\label{eq_rhoMN_inf_MIXaPD}
\end{align}
where we chose the notation $\bar{\rho}_\myCARRIER^\text{t}$ for the total number density of all carriers to indicate that it is an average of $\rhomyCARRIER(\ell)$.
Moreover, Eq.~\eqref{eq_rhoMN_inf_MIXc} becomes
\begin{equation} \label{eq_rhotBEnnn}
\rhomyCARGO=\rhoB +\frac{1}{\ell^*}\int_0^{\ell^*}\mathrm{d}\ell\,\rho_\circledast(\ell)
\end{equation}
in the polydisperse continuum.

\end{document}